\documentclass[12pt, draftclsnofoot, onecolumn]{IEEEtran}
\ifCLASSINFOpdf

\else

\fi
\usepackage[cmex10]{amsmath}
\usepackage{amssymb}
\usepackage{cite}
\usepackage{graphicx}
\usepackage{array,color}
\usepackage{amsmath}
\usepackage{stfloats}
\usepackage{graphicx}
\usepackage{subfigure}
\usepackage{tabularx}
\usepackage{epsfig,epsf,color,balance,cite}
\usepackage{algorithmic}
\usepackage{algorithm}
\usepackage{bm}
\usepackage{textcomp}

\begin{document}

\title{Robust Beamforming Design for Ultra-dense User-Centric C-RAN in the Face of Realistic Pilot Contamination and Limited Feedback}
\author{ Cunhua Pan, Hong Ren, Maged Elkashlan, Arumugam Nallanathan, \IEEEmembership{Fellow, IEEE} and Lajos Hanzo, \IEEEmembership{Fellow, IEEE}
\thanks{C. Pan, M. Elkashlan and A. Nallanathan are with the Queen Mary University of London, London E1 4NS, U.K. (Email:\{c.pan, maged.elkashlan\}@qmul.ac.uk). H. Ren was with National Mobile Communications Research Laboratory, Southeast University, Nanjing 210096, China. She is now with the Queen Mary University of London, London E1 4NS, U.K. (e-mail: renhong@seu.edu.cn). L. Hanzo is with the School of Electronics and Computer Science, University of Southampton, Southampton, SO17 1BJ, U.K. (e-mail:lh@ecs.soton.ac.uk). }
}

\maketitle
\vspace{-1.9cm}
\begin{abstract}
The ultra-dense cloud radio access network (UD-CRAN), in which remote radio heads (RRHs) are densely deployed in the network, is considered. To reduce the channel estimation overhead,  we focus on the design of robust transmit beamforming for user-centric frequency division duplex (FDD) UD-CRANs, where only limited channel state information (CSI) is available. Specifically, we conceive a complete procedure for acquiring the CSI that includes two key steps: channel estimation and channel quantization. The phase ambiguity (PA) is also quantized for coherent cooperative transmission.  Based on the imperfect CSI, we aim for optimizing the  beamforming vectors in order to minimize the total transmit power subject to users'  rate requirements and fronthaul capacity constraints. We derive the closed-form expression of the achievable data rate by exploiting the statistical properties of multiple uncertain terms. Then, we propose a low-complexity iterative algorithm for solving this problem based on the successive convex approximation technique. In each iteration, the Lagrange dual decomposition method is employed for obtaining the optimal beamforming vector. Furthermore, a pair of low-complexity user selection algorithms are provided to guarantee the feasibility of the problem. Simulation results confirm the accuracy  of our robust algorithm in terms of meeting the rate requirements. Finally, our simulation results  verify that using a single bit for quantizing the PA is capable of achieving good performance.
\end{abstract}

\IEEEpeerreviewmaketitle
\section{Introduction}

Ultra dense networks (UDNs), where more and more small base stations (BSs) are deployed within a given area, have been widely regarded as one of the most promising techniques of achieving a high system throughput \cite{Andrews2014}. In UDNs, the average distance between small BSs and users can be dramatically reduced, which can translate into improved link reliability. However, since all small BSs reuse the same frequency, the users are also exposed to severe inter-cell interference, which is a severe performance limiting factor.  Hence, the interference should be judiciously managed in order to reap the potential benefits of UDNs. As a result, the cloud radio access network (CRAN) concept has been recently proposed as a promising network architecture \cite{mugen2016}.
\begin{figure}
\centering
\includegraphics[width=2.6in]{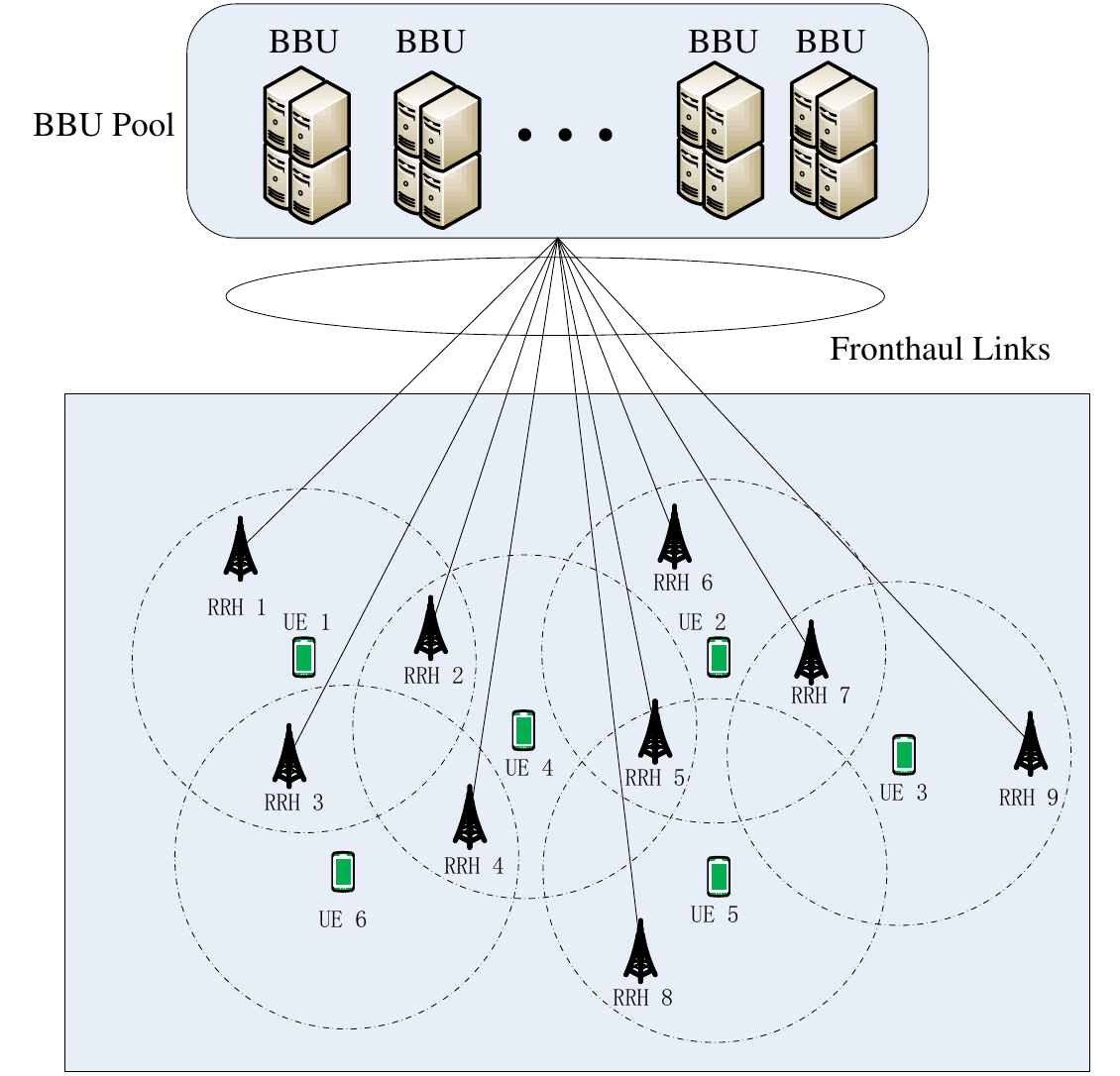}\vspace{-0.5cm}
\caption{Illustration of a UD-CRAN with nine RRHs and six UEs, i.e., $I=9$, $K=6$. To reduce the complexity, each UE is served by the RRHs within the dashed circle centered around the UE. }
\label{fig1}\vspace{-0.8cm}
\end{figure}
In CRAN, all the signal processing tasks are performed at the BBU pool, and all the conventional small BSs are replaced by low-cost low-power RRHs, which are only responsible for simple transmission/reception functions. The RRHs are connected to the BBU pool through the fronthaul links to support the centralized signal processing. Hence, the interference in the network can be effectively mitigated by employing the coordinated multipoint (CoMP) technique. Furthermore, due to their  low-complexity functionalities, the mobile operator can densely deploy the RRHs at a low capital cost. Hence, the CRAN architecture is an ideal platform for supporting UDNs. This kind of network is generally termed as an  UD-CRAN \cite{shi2015IWC,Stephen2017}. An simple example of UD-CRAN is illustrated in Fig.~\ref{fig1}, where the number of RRHs is larger than that of the UEs.

Most of the existing contributions tend to deal with the various technical issues of  conventional CRAN with a limited number of RRHs based on the assumption of the availability of perfect CSI \cite{Yuanming2014,dai2016energy,Binbin2014,vnhatvt2016,Abdelnasser2016,pan2017twc,Luong2016,Luong2017tsp}. In particular, Luong \emph{et al.} \cite{Luong2016} considered the transmit power minimization problem for the downlink of C-RANs with limited fronthaul capacity, where a pair of novel iterative algorithms were proposed for solving this problem. In the first one, the classic successive convex approximation framework was adopted for approximating the continuous nonconvex constraints, and the problem was converted into a mixed-integer second order cone program (MI-SOCP). By relaxing the binary variables to continuous vlaued variables, the second algorithm that is based on the so-called inflation procedure was proposed, which only has to solve a series of SOCP problems. Most recently, the same authors studied in \cite{Luong2017tsp} considered the tradeoff between the achievable sum-rate and total power consumption by using the radical multiobjective optimization concept, where the optimization problem was formulated as a mixed-integer nonconvex program. The authors proposed a branch and reduce and bound-based (BRB) algorithm for finding the globally optimal solution for benchmarking purposes,  and also provided low-complexity iterative algorithms similar to the ones in \cite{Luong2016}.

However, the most challenging issue in UD-CRANs is that a large amount of CSI is required  for facilitating CoMP transmission. The acquisition of the CSI requires a large amount of training resources that escalate rapidly with the network size. One of the most promising solutions  is to consider the availability of only partial CSI. Specifically, each user only has to estimate the CSI of the links from the RRHs in its serving cluster (termed intra-cluster CSI), while only measuring the large-scale channel gains (such as path loss and shadowing) for the CSI of the links from the RRHs beyond its serving cluster (termed inter-cluster CSI). For the example in Fig.~\ref{fig1}, UE 1 only needs to estimate the CSI from RRH 1,2, and 3 to itself, while only the large-scale channel gains are required for the RRHs outside of its cluster. For this kind of scenario, the methods developed in \cite{Yuanming2014,dai2016energy,Binbin2014,vnhatvt2016,Abdelnasser2016,pan2017twc} based on the assumption of perfect CSI cannot be tailored for this case.

Recently, the transmission design relying on  partial CSI  has attracted extensive research interests \cite{Shi2014ICC,Lakshmana2016,Fan2016,pan2017joint}. In particular, a novel compressive CSI acquisition method was proposed in \cite{Shi2014ICC} that can adaptively determine the set of instantaneous CSIs that should be estimated.  The weighted sum-rate maximization problem was considered in \cite{Lakshmana2016}, where the Cauchy-Schwarz inequality was employed for deriving the lower-bound of the accurate data rate. The threshold-based channel matrix sparsification method was proposed in \cite{Fan2016} for a UD-CRAN, where the authors demonstrated that only a negligible performance loss will be caused by discarding the channel matrix entries below a certain threshold.  Finally, in our recent work \cite{pan2017joint}, we proposed a unified framework to deal with the challenges arising in UD-CRAN, and Jensen's inequality was utilized to obtain a more tight lower bound on the achievable rate than that in \cite{Lakshmana2016}.

However, in \cite{Shi2014ICC,Lakshmana2016,Fan2016,pan2017joint}, perfect intra-cluster CSI was assumed to be  available at the BBU pool, which is unrealistic for UD-CRANs, especially when the network operates in the frequency division duplex (FDD) mode \cite{bai2013evolved}, which is the focus of this paper. Tran \emph{et al.} \cite{Tran2106con} considered the queue-aware robust beamforming design to minimize the average transmission power in the face of imperfect CSI for the whole C-RAN, while satisfying the outage probability constraint of each user. The classic Lyapunov optimization theory was employed for ensuring the system's stability. The Bernstein-Type Inequality \cite{lau} was utilized for transforming the outage probability constraints into a more tractable form that facilitates the application of the Semi-Definite Relaxation (SDR) approach. However, the channel error model is only suitable for the channel estimation error.  In FDD UD-CRAN, each user has to estimate the intra-cluster CSI based on the pilot sequences sent from the RRHs within  the serving cluster. Then, the user selects a codeword from a pre-designed CSI codebook to quantize the estimated CSI  and feeds back its index to the BBU pool through a dedicated feedback channel. This procedure will impose three kinds of channel imperfections: channel estimation error, CSI  quantization error and feedback delay. Since UD-CRANs are usually deployed in a limited area such as shopping malls and stadiums where the users move slowly, the effect of channel feedback delay can be ignored \cite{pan2017joint}. However, the other two error sources are inevitable and remain to be a serious problem in UD-CRANs.

To estimate the intra-cluster CSI, the pilot sequences sent from the RRHs that belong to the same user's serving cluster should be mutually orthogonal so that the user can differentiate the channels associated with different RRHs. For the example in Fig.~\ref{fig1}, since RRH 1, 2, and 3 cooperatively serve UE 1, the pilot sequences sent from these RRHs should be mutually orthogonal. A direct method is to assign to all the RRHs  mutually orthogonal pilots. However, the number of pilots linearly increases with the number of RRHs, which is excessive in UD-CRANs. To save the pilot resources, one should allow the RRHs serving no common user to reuse the same pilot. The authors \cite{Chen2016tvt,junzhang2017twc} provided novel pilot reuse schemes for minimizing the total number of pilots required based on graph theory. In \cite{Nguyen2015}, Nguyen \emph{et al.} proposed an iterative pilot allocation method for multicell massive MIMO networks, where the modified Hungarian method was adopted to solve the pilot allocation problem for each cell by fixing the pilot assignments for all the other cells. However, the beamforming direction was fixed and the computational complexity of the pilot assignment algorithm increases drastically with the number of cells. It is commonly known that the pilot reuse scheme will impose non-negligible pilot contamination, which inevitably leads to sizeable channel estimation error that cannot be eliminated. Hence, the channel estimation error should be taken into account when designing the transmission strategy. A robust beamforming design explicitly considering the channel estimation error was studied in our recent work \cite{CunhuTWC} for time division duplex (TDD) UD-CRANs, where no channel quantization error is imposed as a benefit of the TDD channel's reciprocity.

Since coherent cooperative transmission among RRHs provides higher spectral efficiency than non-coherent transmission, we consider the limited feedback scenario of the former transmission scheme. To reduce the implementation complexity, the authors in \cite{disu2011,fangyuantcom} advocated the per-RRH limited feedback strategy, where the estimated channels of all the links from all the candidate RRHs to each user are independently quantized rather than quantizing them jointly. However, this feedback strategy will result in phase ambiguity (PA) \cite{disu2011}. To elaborate, the PA is the phase differences between the single-RRH channel direction information (CDI) and the single-RRH  quantized CSI codeword, which has no impact on the conventional single-cell channel quantization. However, it was shown in \cite{disu2011} that its adverse effect can be compensated by feeding back the PA information to the transmitter at the cost of a modest feedback overhead.

In this paper, we consider the robust downlink beamforming design of  FDD UD-CRAN by taking into account all the channel uncertainties. Specifically, we aim for jointly optimizing the user-RRH associations and beamforming vectors for minimizing the total transmission power subject to  users' rate requirements, fronthaul capacity constraints and per-RRH power constraints.
This is a mixed integer non-linear programming (MINLP) problem that is generally difficult to solve. For the imperfect CSI considered in this paper, in contrast to the  constraints of (6) and (7) in \cite{Luong2016}, the SINR constraints cannot be transformed into an SOCP format. Due to the same reason, the BRB algorithm in \cite{Luong2017tsp} aiming for globally optimal solution cannot be used for the imperfect CSI case. Furthermore, for the low-complexity algorithms developed in \cite{Luong2016,Luong2017tsp}, one has to solve an MI-SOCP or SOCP problem in each iteration, which incurs high computational complexity for UD-CRAN. Specifically, the contributions of this paper are summarized as follows:
\begin{enumerate}
  \item We provide a complete and practical procedure for the BBU pool to acquire the CSI required for centralized signal processing, namely for both channel estimation and channel quantization. To the best of our knowledge, this paper is the first attempt to unify these two steps into a joint framework. We derive the closed-form expression of the achievable data rate by exploiting statistical characteristics of the channel estimation error,  the per-RRH CDI, the PA quantization errors and partial inter-cluster CSI.
  \item To address the feasibility issue, we provide a pair of low-complexity user selection algorithms, namely the successive UE deletion method having a complexity order of $O(K)$ and a bisection based search  method having a complexity order of $O({\rm{log}}_2(K))$, where $K$ is the total number of users. Simulation results show that the former algorithm performs better than the latter, and only slightly worse than the exhaustive search based method having an exponentially increasing complexity order of $K$. The performance loss is roughly 8\% in the worst case.
  \item Based on the feasible set of users given by the user selection algorithms, we propose a low-complexity iterative algorithm for solving the power minimization problem. Specifically, the non-smooth indicator function is approximated as a smooth concave real-valued fractional function, which is iteratively approximated by its first order Taylor expansion. In contrast to \cite{CunhuTWC}, this paper additionally considers the impact of CSI quantization errors, hence the semi-definite relaxation approach developed in \cite{CunhuTWC} cannot be guaranteed to generate a rank-one beamforming solution. Instead, we approximate the complex-valued useful signal part in the rate expression by its first-order Taylor expansion with the aid of the $\cal T$-transform \cite{telatar1999capacity} that transforms complex-valued matrices and vectors into their real-valued equivalents. The transformed optimization problem becomes a convex one, and we derive the optimal beamforming vectors by employing the Lagrange dual decomposition method. Then, the successive convex approximation (SCA) technique is used for iteratively updating the corresponding variables that can guarantee to converge.  Note that \cite{Luong2016,Luong2017tsp} provided the results of the first-order Taylor expansion for the complex-valued expressions without a strict proof. Furthermore, The special structure of the resultant sub-problem has not been exploited for  developing a reduced-complexity algorithm for avoiding the direct solution of the MI-SOCP or SOCP.
\end{enumerate}

The rest of this paper is organized as follows. Section \ref{system} presents the system model. Section \ref{proformu} formulates a two-stage optimization problem. A low-complexity iterative algorithm is provided in Section \ref{powerminimization} to deal with the transmit power minimization problem when the users are selected to be admitted. Two low-complexity user selection algorithms are presented in Section \ref{userselec}. Extensive simulation results are given in Section \ref{simlresult}. Finally, our conclusions are drawn in Section \ref{conclu}.

\emph{Notations}:   ${{\mathbb{E}}_{\{ x\} }}\{ y\} $ denotes the expectation of $y$ over random variable $x$. ${\cal C}{\cal N}(\bf{x}, \bf{\Sigma} )$  denotes the complex Gaussian distribution with mean $\bf{x}$ and variance $\bf{\Sigma}$. The complex set is denoted as ${\mathbb{ C}}$. ${\bf{I}}$ and $\bf{0}$ are an identity matrix and a zero matrix, respectively.  The transpose, conjugate transpose and the pseudo-inverse of matrix $\bf{A}$ are denoted as ${{\bf{A}}^{\rm{T}}}$, ${{\bf{A}}^{\rm{H}}}$ and ${{\bf{A}}^\dag }$, respectively. ${\bf{B}} = {\rm{blkdiag}}\left\{ {{{\bf{A}}_i},i \in {\cal I}} \right\}$ means that matrix ${\bf{B}}$ is formed by performing the block diagonalization over ${{\bf{A}}_i}$. ${\mathop{\rm Re}\nolimits} (\cdot)$ and ${\mathop{\rm Im}\nolimits} (\cdot)$ represent real and imaginary parts of a variable, respectively. $f_{\theta }^\prime (x)$ denotes the first-order derivative of $f_{\theta }(x)$. The other notations are summarized in Table \ref{tabzero}.

\begin{table}[!t]
\renewcommand{\arraystretch}{1.1}
\caption{ The List of Notations}\vspace{-0.5cm}
\label{tabzero}
\centering
\begin{tabular}{|c|c|c|c|}
\hline
$I$  & The number of RRHs   & $K$ & The number of UEs   \\
\hline
$\cal I$ &  The set of RRHs & ${\overline {\cal U}}$ & The set of UEs \\
\hline
$\cal U$ &  The set of selected UEs & ${\cal I}_k$ &The candidate set of RRHs serving UE $k$ \\
\hline
${\cal U}_i$ &  The candidate set of UEs served by RRH $i$ & ${\bf{h}}_{i,k}$ & The channel from RRH $i$ to UE $k$ \\
\hline
${\bf w}_{i,k}$ &  The BF vector from RRH $i$ to UE $k$ & $\alpha _{i,k}$ &Large-scale channel gain from RRH $i$ to UE $k$ \\
\hline
${\bf{\bar h}}_{i,k}$ &  Small-scale fading from RRH $i$ to UE $k$ & $\sigma _k^2$ & Noise power at UE $k$\\
\hline
$M$ &  The number of antennas at each RRH & $\tau$ & The number of time slots for training \\
\hline
${\cal Q}$ &  The set of pilot indices & ${\bf{Q  }}$& The orthogonal pilot sequences  \\
\hline
$n_l$ &  The reuse time of pilot $l$ & $n_{\rm{max}}$& The maximum pilot reuse time  \\
\hline
$ {{\bf{\hat h}}_{i,k}}$& The MMSE estimation of channel $ {{\bf{ h}}_{i,k}}$ & ${{\bf{\tilde h}}_{i,k}}$&  Channel direction information of $ {{\bf{\hat h}}_{i,k}}$ \\
\hline
$ p_t$& Pilot power $ {{\bf{ h}}_{i,k}}$ & ${{\bf{\tilde h}}_{i,k}}$& Channel direction information of $ {{\bf{\hat h}}_{i,k}}$\\
\hline
${{\bf{q}}_{i,k}}$& The quantized version of  ${{\bf{\tilde h}}_{i,k}}$ & ${{{\cal C}_{i,k}}}$&  Per-RRH codebook used by UE $k$ \\
\hline
${B_{i,k}^{{\rm{CDI}}}}$& The number of bits to quantize CDI & $\phi _{i,k}$ & The PA between CDI and its quantized codeword \\
\hline
${{\tilde \phi }_{i,k}}$& The PA quantization error  & ${{\hat \phi }_{i,k}}$ & The quantized version of the PA ${\phi _{i,k}}$ \\
\hline
${B_{i,k}^{{\rm{PA}}}}$& The number of bits to quantize PA  & $a_{i,k}$ & The quantization error of the CDI ${{{\bf{\tilde h}}}_{i,k}}$\\
\hline
\end{tabular}\vspace{-1cm}
\end{table}

\vspace{-0.2cm}\section{System Model}\label{system}
\subsection{Signal Transmission Model}

Consider a downlink  FDD UD-CRAN shown in Fig.~\ref{fig1}, which has $I$ RRHs and $K$ UEs. Each RRH is equipped with $M$ transmit antennas and each UE has a single receive antenna. The sets of RRHs and UEs are denoted as ${\cal I}=\{1,\cdots,I\}$ and ${\overline {\cal U}}=\{1,\cdots,K\}$, respectively. Each RRH is connected to the BBU pool through the wired/wireless fronthaul links. Let ${ {\cal U}} \subseteq \overline {\cal U}$  represent the subset of UEs that can be admitted by the system. To reduce the computational complexity associated with the UD-CRAN, the user-centric cluster technique is considered, where each UE is exclusively served by its nearby RRHs, since the signals arriving from distant RRHs are weak at the UE due to the severe path loss. For the example of Fig.~\ref{fig1}, UE 1 is only potentially served by RRH 1, RRH 2 and RRH 3. The set of RRHs that potentially serve UE $k$ is denoted as ${\cal I}_k\subseteq {\cal I}$, or equivalently the candidate set of RRHs that serve UE $k$ is denoted as ${\cal I}_k$. It should be emphasized that the set of RRHs that finally serve UE $k$ may not be the same as ${\cal I}_k$, which needs to be optimized in the following sections, while the RRHs out of its cluster, i.e.,   ${\cal I}\backslash {\cal I}_k$, will not serve UE $k$.  Additionally, let us denote ${\cal U}_i\subseteq {\cal U}$ as the set of UEs that are potentially served by RRH $i$. Note that the clusters for the UEs may overlap with each other, which means that each RRH can simultaneously serve multiple UEs. These clusters are assumed to be predetermined based on the large-scale channel gains that vary slowly.

Let us denote by ${\bf{h}}_{i,k} \in {{\mathbb{ C}}^{M \times 1}}$ and ${\bf{w}}_{i,k} \in {{\mathbb{ C}}^{M \times 1}}$ the channel vector and beamforming vector of the links spanning from  RRH $i$ to UE $k$, respectively. Then, the  signal received at UE $k$ is
\begin{equation}\label{receivedsignal}
  y_k = \underbrace {\sum\nolimits_{i \in {{\cal I}_k}} {{\bf{h}}_{i,k}^{\rm{H}}{\bf{w}}_{i,k}s_k} }_{{\rm{desired\  signal}}} + \underbrace {\sum\nolimits_{l \ne k,l \in {\cal U}} {\sum\nolimits_{i \in {{\cal I}_l}} {{\bf{h}}_{i,k}^{\rm{H}}{\bf{w}}_{i,l}s_l} } }_{ {\rm{interference}}} + z_k,
\end{equation}
where $s_l$ denotes the transmission data for UE $l$ and $z_k$ is the zero-mean additive complex white Gaussian noise with variance $\sigma _k^2$. It is assumed that the data destined for each UE is independent of each other and it has a zero mean and unit variance, i.e., we have ${{\mathbb{E}}}\{ |s_k{|^2}\}  = 1$ and ${{\mathbb{E}}}\{ s_{{k_1}}s_{{k_2}}\}  = 0$ for $k_1 \ne k_2, \forall k_1, k_2 \in {\cal U}$. The channel vector ${\bf{h}}_{i,k}$ can be decomposed as ${\bf{h}}_{i,k} = \sqrt {\alpha _{i,k}} {\bf{\bar h}}_{i,k}$, where $\alpha _{i,k}$ represents the large-scale channel gains of the links spanning from RRH $i$ to UE $k$ that accounts both for the shadowing and path loss,  while ${\bf{\bar h}}_{i,k}$ is the small-scale channel fading with the distribution of ${\cal C}{\cal N}({\bf{0}},{\bf{I}})$.

\vspace{-0.5cm}\subsection{Channel Estimation for Intra-cluster CSI}\label{chanesti}

To design the beam-vectors for the UEs, the overall CSI should be available at the BBU pool for the facilitation of joint transmission. However, it is an unaffordable task to estimate the CSI from all RRHs to all UEs due to the limited availability of training resources. An appealing approach is  that each UE only estimates the CSI within its cluster, named intra-cluster CSI. For the CSI beyond this cluster, it is assumed that only large-scale channel gains are available, i.e., $\{\alpha _{i,k}, \forall i\in {\cal I}\backslash {{\cal I}_k},k\in {\cal U}\}$. The out-cluster large-scale channel gains are used to control the multiuser interference.

In this paper, we assume that $\tau$ time slots are used for CSI training, thus the length of pilot sequences is $\tau$, or equivalently the number of orthogonal pilot sequences is equal to $\tau$. Let us denote the set of pilot indices as ${\cal Q}=\{1,2,\cdots,\tau\}$, and the corresponding orthogonal pilot sequences as ${\bf{Q = }}\left[ {{{\bf{q}}_1}, \cdots ,{{\bf{q}}_\tau }} \right] \in {\mathbb{ C}^{\tau  \times \tau }}$ that satisfies the orthogonal condition  ${{\bf{Q}}^H}{\bf{Q}} = {\bf{I}}$.

For the channel estimation in an FDD UD-CRAN system, the RRHs first send the training sequences to the UEs,  then the UEs estimate their channels based on their received signals. Specifically, the training signals received at UE $k$ can be written as
\begin{equation}\label{pilotrece}
{{\bf{y}}_k} = \sum\nolimits_{i \in {{\cal I}_k}} {\sqrt {{p_t}} {\bf{h}}_{i,k}^{\rm{H}}} {{\bf{X}}_i^{\rm{H}}} + \sum\nolimits_{i \in {\cal I}/{\cal I}{_k}} {\sqrt {{p_t}} {\bf{h}}_{i,k}^{\rm{H}}} {{\bf{X}}_i^{\rm{H}}} + {{\bf{n}}_k},
\end{equation}
where $p_t$ is the pilot transmit power at each transmit antenna, ${{\bf{n}}_k} \in {{\mathbb C}^{1 \times \tau }}$ is the additive Gaussian noise vector during the training time slots, whose elements are independently generated and follow the distributions of ${\cal C}{\cal N}(0, \sigma _k^2)$, ${{\bf{X}}_i} \in {{\mathbb C}^{\tau  \times M}}$ is the pilot training matrix sent from RRH $i$. The training matrix ${{\bf{X}}_i}$ can be written as ${{\bf{X}}_i} = \left[ {{{\bf{q}}_{\pi _i^1}}, \cdots ,{{\bf{q}}_{\pi _i^M}}} \right]$, where ${{\bf{q}}_{\pi _i^m}}\in {\mathbb{ C}^{\tau  \times 1 }}$ denotes the pilot sequence used for estimating the channels spanning from the $m$th antenna of RRH $i$ to the UEs.

To conserve the pilot resources, a pilot reuse scheme is considered, which should satisfy the following constraints: 1) The pilot sequences from different RRHs in the same cluster should also be orthogonal, i.e. ${\bf{X}}_m^{\rm{H}}{{\bf{X}}_n} = {\bf{0}}$ for $m,n \in {\cal I}{_k},m \ne n, \forall k\in \cal U$; 2) The maximum reuse time for each pilot sequence should be restricted to a small value for reducing the channel estimation error. Let us denote the reuse time for pilot $l$ as $n_l$. Then this condition can be expressed as $n_l\leq n_{\rm{max}},\forall l\in {\cal Q}$. 3) The pilot sequences used by all antennas at the same RRH should be mutually orthogonal, i.e. ${\bf{X}}_i^{\rm{H}}{{\bf{X}}_i} = {\bf{I}}$.  The first constraint means that the RRHs serving the same UE should use an orthogonal pilot matrix. A natural pilot allocation approach to satisfy the above three constraints is the orthogonal pilot  allocation scheme, where all antennas and RRHs are allocated orthogonal pilots. Obviously, the number of pilots required  is $MI$, which occupies lots of time slots for UD-CRANs having a large number of RRHs. Hence, if we allow some RRHs to reuse the same set of pilots, the number of pilot sequences required will be reduced. In this paper, we aim for minimizing the number of pilots required, while guaranteeing the above three conditions. This pilot allocation problem has been studied in \cite{Chen2016tvt}, where the Dsatur algorithm from graph theory was proposed to solve it. The computational complexity of the Dsatur algorithm is given by ${\cal O}\left( {{I^2}} \right)$ \cite{Chen2016tvt}. When some RRHs are allocated the same color, these RRHs can reuse the same pilot. Denote $c^{\star}$ as the number of different colors required by the Dsatur algorithm to color all the RRHs. Then the total number of pilots required is given by $\tau=Mc^{\star}$, since the antennas in each RRH use different pilots.

Let us define ${{\cal K}_{\bf{X}}}{\rm{ = }}\left\{ {i:{{\bf{X}}_i} = {\bf{X}}} \right\}$ as the set of RRHs that reuse the same pilots ${\bf{X}}$ obtained by using the Dsatur algorithm. Then, the MMSE estimation of channel ${\bf{h}}_{i,k}$ is given by \cite{kailath2000linear}
\begin{equation}\label{dewae}
 {{\bf{\hat h}}_{i,k}} = \frac{{{\alpha _{i,k}}}}{{\sum\nolimits_{m \in {{\cal K}_{{{\bf{X}}_i}}}} {{\alpha _{m,k}}}  + {{\hat \sigma }^2}}}\frac{1}{{\sqrt {{p_t}} }}{\bf{X}}_i^{\rm{H}}{\bf{y}}_k^{\rm{H}},
\end{equation}
where ${{\hat \sigma }^2} = {\sigma _k^2}/p_t$. It can be readily derived from (\ref{dewae}) that the channel estimate ${{\bf{\hat h}}_{i,k}}$ obeys the distribution of ${\cal C}{\cal N}({\bf{0}},{\omega_{i,k}}{\bf{I}})$ with $\omega_{i,k}$ given by
\begin{equation}\label{deafre}
 {\omega _{i,k}} = \frac{{\alpha _{i,k}^2}}{{\sum\nolimits_{m \in {{\cal K}_{{{\bf{X}}_i}}}} {{\alpha _{m,k}}}  + {{\hat \sigma }^2}}}.
\end{equation}
According to the property of MMSE estimation \cite{kailath2000linear}, the channel estimate error ${{\bf{e}}_{i,k}} = {{\bf{h}}_{i,k}} - {{{\bf{\hat h}}}_{i,k}}$  is independent of the channel estimate ${{\bf{\hat h}}_{i,k}}$, which follows the distribution of ${\cal C}{\cal N}({\bf{0}},{\delta _{i,k}}{\bf{I}})$, with ${\delta _{i,k}}$ given by
\begin{equation}\label{degrag}
{\delta_{i,k}} = \frac{{{\alpha _{i,k}}\left( {\sum\nolimits_{m \in {{\cal K}_{{{\bf{X}}_i}}}\backslash i} {{\alpha _{m,k}}}  + {{\hat \sigma }^2}} \right)}}{{\sum\nolimits_{m \in {{\cal K}_{{{\bf{X}}_i}}}} {{\alpha _{m,k}}}  + {{\hat \sigma }^2}}}.
\end{equation}
Note that even when RRH $i$ does not reuse any pilots of any other RRHs, there is still some channel estimation error for channel ${{\bf{h}}_{i,k}}$ with ${\delta _{i,k}} = {{{\alpha _{i,k}}{{\hat \sigma }^2}} \mathord{\left/
 {\vphantom {{{\alpha _{i,k}}{{\hat \sigma }^2}} {\left( {{\alpha _{i,k}} + {{\hat \sigma }^2}} \right)}}} \right.
 \kern-\nulldelimiterspace} {\left( {{\alpha _{i,k}} + {{\hat \sigma }^2}} \right)}}$.

\vspace{-0.5cm}\subsection{Limited Feedback Model}\label{limitfeed}

In this paper, we consider the limited  per-RRH codebook feedback strategy \cite{disu2011}, where each UE uses different codebooks to independently quantize its per-RRH CDI, i.e., ${{\bf{\tilde h}}_{i,k}}{\rm{ = }}{{{{{\bf{\hat h}}}_{i,k}}} \mathord{\left/
 {\vphantom {{{{{\bf{\hat h}}}_{i,k}}} {\left\| {{{{\bf{\hat h}}}_{i,k}}} \right\|}}} \right.
 \kern-\nulldelimiterspace} {\left\| {{{{\bf{\hat h}}}_{i,k}}} \right\|}}$. Then UE $k$ feeds back the indices of codewords to its corresponding serving RRHs. The BBU pool will collect all the indices sent from different RRHs and will design  beamforming vectors based on these indices. Specifically, the quantized version of the CDI ${{\bf{\tilde h}}_{i,k}}$ is given by
 \begin{equation}\label{grs}
   {{\bf{q}}_{i,k}} = \mathop {\arg \max }\limits_{{{\bf{c}}_{i,k,n}} \in {{\cal C}_{i,k}}} \left| {{{{\bf{\tilde h}}}_{i,k}^{\rm{H}}}{{\bf{c}}_{i,k,n}}} \right|,
 \end{equation}
where ${{{\cal C}_{i,k}}}$ is the per-RRH codebook used by UE $k$ to quantize the CSI spanning from RRH $i$, which consists of unit-norm codewords ${{\bf{c}}_{i,k,n}}\in {{\mathbb{C}}^{M \times 1}},n = 1, \cdots, {2^{B_{i,k}^{{\rm{CDI}}}}}$, with $B_{i,k}^{{\rm{CDI}}}$ denoting the number of bits used for quantizing the CDI ${{\bf{\tilde h}}_{i,k}}$.

Coherent joint transmission is assumed in this paper. Then, another important parameter namely the phase ambiguity (PA) is also required at the BBU pool \cite{disu2011,fangyuantcom}. The PA is defined as the angle between the per-RRH CDI and its quantized codeword, i.e., ${e^{j{\phi _{i,k}}}} = {{{\bf{\tilde h}}_{i,k}^{\rm{H}}{{\bf{q}}_{i,k}}} \mathord{\left/
 {\vphantom {{{\bf{\tilde h}}_{i,k}^{\rm{H}}{{\bf{q}}_{i,k}}} {\left\| {{\bf{\tilde h}}_{i,k}^{\rm{H}}{{\bf{q}}_{i,k}}} \right\|}}} \right.
 \kern-\nulldelimiterspace} {\left\| {{\bf{\tilde h}}_{i,k}^{\rm{H}}{{\bf{q}}_{i,k}}} \right\|}}$ with $j = \sqrt { - 1} $. The PA knowledge is not required for single-point limited feedback MIMO systems, but affects the co-phasing of the coherent joint transmission in UD-CRAN, as detailed in \cite{disu2011,fangyuantcom}.
The PA can be fed back with the aid of a few bits by using scalar quantization. Since the codeword is chosen by maximizing the magnitude of ${{{{\bf{\tilde h}}}_{i,k}}{{\bf{c}}_{i,k,n}}}$ and the CDI is isotropically distributed, the PA ${\phi _{i,k}}$ will be uniformly distributed in $\left[ {0,2\pi } \right]$. Hence, it is optimal to quantize the PA employing a uniform scalar quantizer. Let us denote by ${{\tilde \phi }_{i,k}}$ and ${{\hat \phi }_{i,k}}$ the PA quantization error and the quantized version of the PA ${\phi _{i,k}}$, respectively. Then, the PA ${\phi _{i,k}}$ can be represented as ${\phi _{i,k}} = {{\hat \phi }_{i,k}} + {{\tilde \phi }_{i,k}}$. If we use $B_{i,k}^{{\rm{PA}}}$  bits to quantize PA ${\phi _{i,k}}$, the PA quantization error ${{\tilde \phi }_{i,k}}$ is uniformly distributed within $\left[ { - \frac{\pi }{{{2^{B_{i,k}^{{\rm{PA}}}}}}},\frac{\pi }{{{2^{B_{i,k}^{{\rm{PA}}}}}}}} \right]$.

Let us define by $a_{i,k} \buildrel \Delta \over = 1 - {\left| {{{{\bf{\tilde h}}}_{i,k}}{{\bf{q}}_{i,k}}} \right|^2}$  the quantization error of the CDI ${{{\bf{\tilde h}}}_{i,k}}$. For simplicity, random vector quantization (RVQ) is considered for quantizing the per-RRH CDIs in this paper. Then, according to \cite{Jinadl}, the per-RRH CDI ${{\bf{\tilde h}}_{i,k}}$ can be rewritten as
\begin{equation}\label{deafe}
 {{{\bf{\tilde h}}}_{i,k}} = \sqrt {1 - a_{i,k}} {e^{j{\phi _{i,k}}}}{{\bf{q}}_{i,k}} + \sqrt {a_{i,k}} {{\bf{u}}_{i,k}},
\end{equation}
where ${{\bf{u}}_{i,k}}$ is channel quantization error, which is a unit-norm vector isotropically distributed in the nullspace of ${{\bf{q}}_{i,k}}$.

In this paper, we assume that there are dedicated error-free feedback channels for feeding back all quantized versions of CDIs and PAs to the BBU pool. Then, the BBU pool determines the  beamforming vectors based on the feedback information.

\vspace{-0.5cm}\section{Problem Formulation}\label{proformu}

In this section, we first provide the mathematical model for the constraints of the UD-CRAN, which include each UE's data rate requirement, the per-RRH power constraint and limited fronthaul capacity constraint. Then, based on these constraints, we formulate the UE selection problem and the transmit power minimization problem in a two-stage form.

Let us denote the beamforming vectors from all RRHs in ${\cal I}_k$ by $ { {{\bf{ w}}}}_k = {[ {{\bf{ w}}_{i,k}^{{\rm{H}}},\forall i \in {{\cal I}_k}} ]^{\rm{H}}} \in {{\mathbb{C}}^{\left| {{{\cal I}_k}} \right|M \times 1}}$, and the aggregated channel vectors from  RRHs in ${\cal I}_l$ to UE $k$ by ${\bf{g}}_{l,k}= [ {{\bf{h}}_{i,k}^{{\rm{H}}},\forall i \in {{\cal I}_l}} ]^{{\rm{H}}} \in {{\mathbb C}^{\left| {{{\cal I}_l}} \right|M \times 1 }}$. In addition, define ${\bf{\tilde g}}_{k,k}= [{\bf{e}}_{i,k}^{\rm{H}},\forall i \in {{\cal I}_k}]^{\rm{H}}  \in {{\mathbb C}^{\left| {{{\cal I}_k}} \right|M \times  1}}$ and ${\bf{ \hat g}}_{k,k} = [{\bf{\hat h}}_{i,k}^{\rm{H}},\forall i \in {{\cal I}_k}]^{\rm{H}}\in {{\mathbb C}^{\left| {{{\cal I}_k}} \right|M \times 1}}$ as the overall CSI error and estimated CSI of the links spanning from the RRHs in ${\cal I}_k$ to UE $k$, respectively. Then, the channel estimation error can be rewritten as ${{{\bf{\tilde g}}}_{k,k}} = {{\bf{g}}_{k,k}} - {{{\bf{\hat g}}}_{k,k}}$, while the received signal model in (\ref{receivedsignal}) can be reformulated as
\begin{equation}\label{achieveablerate}
  {y_k} = \underbrace {{\bf{\hat g}}_{k,k}^{\rm{H}}{{\bf{w}}_k}{s_k}}_{{\rm{Desired\  signal}}} + \underbrace {{\bf{\tilde g}}_{k,k}^{\rm{H}}{{\bf{w}}_k}{s_k}}_{{\rm{Residual- interference}}} + \underbrace {\sum\nolimits_{l \ne k,l \in {\cal U}} {{\bf{g}}_{l,k}^{\rm{H}}{{\bf{w}}_l}{s_l}} }_{{\rm{ Multi-user\  Interference}}} + {z_k},\forall k \in  {\cal U}.
\end{equation}
As in most existing papers \cite{Jose2011,Chien2016}, we consider the achievable data rate, where the residual-interference term in (\ref{achieveablerate}) due to the channel estimation error is treated as uncorrelated Gaussian noise. Additionally, for the sake of reducing the decoding complexity, the multi-user interference term is also regarded as uncorrelated Gaussian noise. By considering the time slots allocated for channel training, the net achievable data rate of UE $k$ can be expressed as \cite{Chien2016}
\begin{equation}\label{datarateUEk}
  {r_k} = \frac{{T - \tau }}{T} {{{\log }_2}\left( {1 + \frac{{\mathbb{E}} \left\{{{{\left| {{\bf{\hat g}}_{k,k}^{\rm{H}}{{\bf{w}}_{k}}} \right|}^2}}\right\}}
  {{{\mathbb{E}}\left\{{{\left| {{\bf{\tilde g}}_{k,k}^{\rm{H}}{{\bf{w}}_k}} \right|}^2}\right\} + \sum\nolimits_{l \ne k,l \in {\cal U}} {{\mathbb{E}}\left\{{{\left| {{\bf{g}}_{l,k}^{\rm{H}}{{\bf{w}}_l}} \right|}^2}\right\} + \sigma _k^2} }}} \right)} , \forall k\in {\cal U},
\end{equation}
where $\tau$ is the total number of time slots required by the Dsatur algorithm,  $T$ denotes the total number of time slots in each time frame, and the expectation is taken over multiple random processes, namely, the fast fading of the unknown CSI in ${\cal I}\backslash { {\cal I}_k}$, the channel estimation errors $\left\{ {{{\bf{e}}_{i,k}},i \in {{\cal I}_k}} \right\}$, the CDI quantization errors $\left\{{{\bf{u}}_{i,k}},\forall i\in {\cal I}_k\right\}$ and the PA quantization errors $\left\{{{\tilde \phi }_{i,k}},\forall i\in {\cal I}_k\right\}$. Each UE's data rate should be higher than its minimum rate requirement:
\begin{equation}\label{hoieowe}
  {\rm{C1:}}\  {r_k}\geq {R_{k,\min }}, \forall k\in \cal U,
\end{equation}
where ${R_{k,\min }}$ is the rate target of UE $k$.

The second constraint is the per-RRH power constraint, which can be expressed as
\begin{equation}\label{powerconst}
{\rm{C2}}:\;\sum\nolimits_{k \in {{\cal U}_i}} {{\left\| {{\bf{w}}_{i,k}} \right\|}^2}  \le {P_{i,\max }},i \in {\cal I},
\end{equation}
where ${P_{i,\max }}$ is the power limit of RRH $i$.

Finally, each fronthaul link has a capacity constraint, since we consider a limited bandwidth. Specifically, this kind of constraint can be expressed as
\begin{equation}\label{fronthaulca}
  {\rm{C3}:}\ \sum\nolimits_{k \in {\cal U}_i} {\varepsilon \left({\left\| {{\bf{w}}_{i,k}} \right\|}^2\right){r}_{k}}  \le {C_{i,\max }},\forall i \in {\cal I},
\end{equation}
where $C_{i,\max }$ is the capacity limit of the fronthaul link spanning from the BBU pool to RRH $i$, and $\varepsilon \left( \cdot \right)$ is an indicator function,  defined as
\begin{equation}\label{indifunc}
  \varepsilon \left( x \right) = \left\{ {\begin{array}{*{20}{l}}
{1,\;{\rm{if}}\;x \ne 0,}\\
{0,\;{\rm{otherwise}}.}
\end{array}} \right.
\end{equation}

Due to the constraints of the system (C2 and C3), some UEs' rate requirements (C1) may not be satisfied. Hence, some UEs should be removed in order to guarantee the QoS requirements of the remaining UEs. Similar to \cite{pan2017twc,pan2017joint}, we formulate a two-stage  optimization.

Specifically, in Stage I, we aim for maximizing the number of UEs admitted to the dense network, which is formulated as
\begin{equation}\label{staone}
\begin{array}{l}
 {\cal P}_1:\ \mathop {\max }\limits_{{\bf{w}}, {\cal U} \subseteq {\overline {\cal U}} } \quad \left| \cal U \right|\\
\qquad\ {\rm{s}}.{\rm{t}}.\qquad {\kern 1pt} {\rm{C1}},{\rm{C2}},{\rm{C3}},
\end{array}
\end{equation}
where $\bf{w}$ denotes the set of all beamforming vectors and $\left| \cal U \right|$ is the cardinality of the set $\cal U $.

In Stage II, our goal is to optimize the beamforming vectors for minimizing the total transmit power with the UEs selected from Stage I. Let us denote by ${{\cal U}^\star}$ the specific solution from Stage I, where the corresponding ${{ {\cal U}}_i}$ becomes ${{ {\cal U}}_i^\star}$. Then the optimization problem in Stage II is
\begin{equation}\label{statwo}
\begin{array}{l}
 {\cal P}_2:\ \mathop {\min }\limits_{{\bf{w}}} \quad \sum\nolimits_{i \in {\cal I}} {\sum\nolimits_{k \in {{\cal U}_i^\star}} {\left\| {{{\bf{w}}_{i,k}}} \right\|_2^2} } \\
\qquad\ {\rm{s}}.{\rm{t}}.\qquad {\kern 1pt} {\rm{C1}},{\rm{C2}},{\rm{C3}}.
\end{array}
\end{equation}
In constraints C1-C3, ${{\cal U}}$ and ${{ {\cal U}}_i}$ are replaced by ${{\cal U}^\star}$ and ${{ {\cal U}}_i^\star}$, respectively.

Problems ${\cal P}_1$ and ${\cal P}_2$ in (\ref{staone}) and (\ref{statwo}) are difficult to solve. The reasons are given as follows. Firstly, the exact data rate ${r_k}$ is difficult to derive, since the expectation is taken over multiple uncertain terms. Secondly, both the objective function and the fronthaul capacity constraint C3 of Problem ${\cal P}_1$ contain the non-smooth and non-differential indicator functions, which is recognized as a mixed-integer non-linear programming (MINLP) problem. The exhaustive search method can be adopted to solve this kind of optimization problem. However, it has an exponential complexity order, which becomes excessive for UD-CRAN with large number of UEs.

In the following section, we first deal with the power minimization Problem ${\cal P}_2$ by assuming that the set of admitted UEs has already been determined by solving Problem ${\cal P}_1$. Then, we will conceive low-complexity methods to deal with Problem ${\cal P}_1$ in Section \ref{userselec}.

\vspace{-0.5cm}\section{Low-complexity Algorithm to Deal with Problem ${\cal P}_2$}\label{powerminimization}

In this section, we provide a low-complexity algorithm for solving Problem ${\cal P}_2$, when the UEs to be admitted have already been selected by using the UE selection algorithms in Section \ref{userselec}, and denote the subset of UEs that have been selected as $\cal U$. In the following, we first simplify the rate expression.

The multiple random processes in the rate expression make the accurate closed-form expression of the achievable data rate of UE $k$ in (\ref{datarateUEk}) difficult to derive. In Appendix \ref{prooftheorem1}, we derived the achievable data rate as
\begin{eqnarray}
 r_k &=& \frac{{T - \tau }}{T}{\log _2}\left( {1 + \frac{{{\bf{w}}_k^{\rm{H}}{{\bf{A}}_{k,k}}{{\bf{w}}_k}}}{{{\bf{w}}_k^{\rm{H}}{{\bf{E}}_{k,k}}{{\bf{w}}_k} + \sum\nolimits_{l \ne k,l \in {\cal U}} {{\bf{w}}_l^{\rm{H}}{{\bf{A}}_{l,k}}{{\bf{w}}_l} + \sigma _k^2} }}} \right)\label{secondg}\\
 &=&\frac{{T - \tau }}{T}{\log _2}\left( {1 + {\rm{SIN}}{{\rm{R}}_k}} \right),\label{flogtsu}
\end{eqnarray}
where we have ${{\bf{E}}_{k,k}} = {\mathbb{E}}\left\{ {{{{\bf{\tilde g}}}_{k,k}}{\bf{\tilde g}}_{k,k}^{\rm{H}}} \right\}\in{\mathbb{C}}^{M{|{\cal I}_k|} \times M{|{\cal I}_k|}}$,
 ${{\bf{A}}_{k,k}} = {\mathbb{E}}\left\{ {{{{\bf{\hat g}}}_{k,k}}{\bf{\hat g}}_{k,k}^{\rm{H}}} \right\}\in {\mathbb{C}}^{M{|{\cal I}_k|} \times M{|{\cal I}_k|}}$ and ${\bf{A}}_{l,k} = {{\mathbb{E}}}\left\{ {{\bf{g}}{{_{l,k}}}{\bf{g}}_{l,k}^{\rm{H}}} \right\}\in {\mathbb{C}}^{M{|{\cal I}_l|} \times M{|{\cal I}_l|}}$. The matrix ${{\bf{E}}_{k,k}}$ can be readily computed as
 \begin{equation}\label{jor}
   {{\bf{E}}_{k,k}} = {\rm{blkdiag}}\left\{ {{\delta _{i,k}}{{\bf{I}}_{M}},i \in {{\cal I}_k}} \right\},
 \end{equation}
 while ${\bf{A}}_{k,k}$  and ${\bf{A}}_{l,k}$ are given in (\ref{akk}) and (\ref{jiofuh}) of Appendix \ref{prooftheorem1}, respectively. Note that the matrices ${{\bf{E}}_{k,k}}$, ${{\bf{A}}_{k,k}}$ and ${\bf{A}}_{l,k} $ are semi-definite matrices, since they represent the expectations over semi-definite matrices \cite{boyd2004convex}. The achievable signal to interference plus noise ratio (SINR) of UE $k$ is given by
\begin{equation}\label{rjlotlo}
 {\rm{SIN}}{{\rm{R}}_k} = \frac{{{\bf{w}}_k^{\rm{H}}{{\bf{A}}_{k,k}}{{\bf{w}}_k}}}{{{\bf{w}}_k^{\rm{H}}{{\bf{E}}_{k,k}}{{\bf{w}}_k} + \sum\nolimits_{l \ne k,l \in {\cal U}} {{\bf{w}}_l^{\rm{H}}{{\bf{A}}_{l,k}}{{\bf{w}}_l} + \sigma _k^2} }}.
\end{equation}

By  exploiting the fact that the rate constraints hold with equality at the optimal point \cite{CunhuTWC}, Problem ${\cal P}_2$ can be transformed as
\begin{subequations}\label{azdfS}
\begin{align}
{\cal P}_3:\ \mathop {\min }\limits_{{\bf{w}} }\quad
& \sum\nolimits_{i \in {\cal I}} {\sum\nolimits_{k \in {{\cal U}_i}} {\left\| {{{\bf{w}}_{i,k}}} \right\|_2^2} } \\
\qquad\ \textrm{s.t.}\qquad\!\!\!\!
&{\rm{C2}}, {\rm{C4:}} \ {\rm{SIN}}{{\rm{R}}_k} \ge {\eta _{k,\min }},\forall i \in {\cal I}, \\
&{\rm{C5:}}\ \sum\nolimits_{k \in {\cal U}_i} {\varepsilon \left( {\left\| {{\bf{w}}_{i,k}} \right\|}^2\right){R_{k,\min }}}  \le {C_{i,\max }},\forall i \in {\cal I},
\end{align}
\end{subequations}
where ${\eta _{k,\min }} = {2^{\frac{T}{{T - \tau }}{R_{k,\min }}}} - 1$.

\vspace{-0.5cm}\subsection{Smooth Approximation of the Indicator Function}\label{sonemo}
We first deal with the non-smooth nature of the indicator function in C5. Similar to \cite{pan2017joint}, the indicator function is approximated by the smooth function ${f_\theta }(x) = \frac{x}{{x + \theta }}$, where $\theta$ is a small constant. By replacing the indicator function  with ${f_\theta }(x)$, Problem ${\cal P}_3$ can be approximated as
\vspace{-0.2cm}
\begin{subequations}\label{appro}
\begin{align}
{\cal P}_4:\ \mathop {\min }\limits_{{\bf{w}} }\quad
& \sum\nolimits_{i \in {\cal I}} {\sum\nolimits_{k \in {{\cal U}_i}} {\left\| {{{\bf{w}}_{i,k}}} \right\|_2^2} } \\
\qquad\ \textrm{s.t.}\qquad\!\!\!\!
&{\rm{C2}}, {\rm{C4}}, {\rm{C6:}}\ \sum\nolimits_{k \in {{\cal U}_i}} {{f_\theta }\left( {{{\left\| {{{\bf{w}}_{i,k}}} \right\|}^2}} \right){R_{k,\min }}}  \le {C_{i,\max }},\forall i \in {\cal I}.
\end{align}
\end{subequations}
The successive convex approximation (SCA) method \cite{dinh2010local} is used to deal with the non-convex constraint C7. Specifically, by exploiting the concavity of ${f_\theta }(x)$, we have
\vspace{-0.2cm}
\begin{equation}\label{fucntions}
 {f_\theta }\left( {{{\left\| {{{\bf{w}}_{i,k}}} \right\|}^2}} \right) \le {f_\theta }\left( {{{\left\| {{{\bf{w}}_{i,k}(t)}} \right\|}^2}} \right) + { \beta }_{i,k}(t) \left( { {{{\left\| {{{\bf{w}}_{i,k}}} \right\|}^2}}  - {{{\left\| {{{\bf{w}}_{i,k}(t)}} \right\|}^2}} } \right),\vspace{-0.2cm}
\end{equation}
where ${\bf{w}}_{i,k}(t)$ is the beamforming vector at the $t$th iteration, $\beta_{i,k}(t) = {f_\theta ^\prime }\left( {{{\left\| {{{\bf{w}}_{i,k}(t)}} \right\|}^2}} \right)$. By replacing ${f_\theta }\left( {{{\left\| {{{\bf{w}}_{i,k}}} \right\|}^2}} \right)$ in Problem ${\cal P}_4$ with the right hand side of (\ref{fucntions}), we arrive at
\vspace{-0.2cm}
\begin{subequations}\label{jeouiruj}
\begin{align}
{\cal P}_5:\ \mathop {\min }\limits_{{\bf{w}} }\quad
& \sum\nolimits_{i \in {\cal I}} {\sum\nolimits_{k \in {{\cal U}_i}} {\left\| {{{\bf{w}}_{i,k}}} \right\|_2^2} } \\
\qquad\ \textrm{s.t.}\qquad\!\!\!\!
&{\rm{C2}}, {\rm{C4}}, {\rm{C7:}}\ \sum\nolimits_{k \in {{\cal U}_i}} {{\tau _{i,k}(t)}{{\left\| {{{\bf{w}}_{i,k}}} \right\|}^2}}  \le {{\tilde C}_{i }(t)}, \forall i \in {\cal I},
\end{align}
\end{subequations}
where ${\tau _{i,k}(t)} = {\beta _{i,k}}(t){R_{k,\min }}$, ${{\tilde C}_i(t)} = {C_{i,\max }} - \sum\nolimits_{k \in {{\cal U}_i}} {\left( {{f_\theta }\left( {{{\left\| {{{\bf{w}}_{i,k}(t)}} \right\|}^2}} \right) - {\beta _{i,k}(t)}{{\left\| {{{\bf{w}}_{i,k}(t)}} \right\|}^2}} \right){R_{k,\min }}} $.

However, Problem ${\cal P}_5$ is still difficult to solve due to Constraint C4, although it has been simplified from Constraint C1.  The reasons are given as follows. Due to the channel estimation error, each user suffers from residual interference, as seen from the right hand side of C4, i.e. ${\bf{w}}_k^{\rm{H}}{{\bf{E}}_{k,k}}{{\bf{w}}_k}$. Although the classic weighted minimum mean square error (WMMSE) method  has been successfully applied in UD-CRANs under the idealized simplifying assumptions of having perfect intra-cluster CSI \cite{pan2017twc,pan2017joint,Binbin2014}, it cannot be adopted in this realistic optimization problem due to the residual-interference. Furthermore, note that the rank of matrix ${{\bf{A}}_{k,k}}$ is in general higher than one, the Semi-definite (SDP) relaxation method used in \cite{CunhuTWC} cannot be adopted here, since the resultant solution is no longer guaranteed to be of rank one. In the following, we propose a novel method to deal with Constraint C4.

\vspace{-0.3cm}\subsection{Method to Deal with Constraint C4}\label{freito}

In this following, we propose a novel method based on the first-order Taylor approximation  to deal with Constraint C4 and then propose the Lagrange dual decomposition algorithm for solving this problem.

Constraint C4 is non-convex, because ${\bf{w}}_k^{\rm{H}}{{\bf{A}}_{k,k}}{{\bf{w}}_k}$ is a convex function of ${{\bf{w}}_k}$\footnote{Note that ${{\bf{A}}_{k,k}}$ is a semi-definite matix.}. Similar to the successive convex approximation method  dealing with the concave fractional function, we approximate it by its first-order Taylor expansion  and make Constraint C4 convex. Since ${\bf{w}}_k^{\rm{H}}{{\bf{A}}_{k,k}}{{\bf{w}}_k}$ is convex, we have
\begin{equation}\label{ehdliu}
 {\bf{w}}_k^{\rm{H}}{{\bf{A}}_{k,k}}{{\bf{w}}_k} \ge {\bf{w}}_k^{\rm{H}}(t){{\bf{A}}_{k,k}}{{\bf{w}}_k}(t) + 2{\mathop{\rm Re}\nolimits} \left\{ {{\bf{w}}_k^{\rm{H}}(t){{\bf{A}}_{k,k}}\left( {{{\bf{w}}_k} - {{\bf{w}}_k}(t)} \right)} \right\},
\end{equation}
where ${{\bf{w}}_k}(t)$ is the beamforming vector at the $t$th iteration. The above derivation is not direct since ${\bf{w}}_k^{\rm{H}}{{\bf{A}}_{k,k}}{{\bf{w}}_k}$ is a function of complex-valued vector ${{\bf{w}}_k}$. The Taylor expansion developed for the functions over real-valued variables cannot be directly extended to the complex case. In Appendix \ref{hfiureah}, we derived the above result relying on the so-called $\cal T$-transform \cite{telatar1999capacity} that transforms complex-valued matrices and vectors into their real-valued equivalents.

By replacing ${\bf{w}}_k^{\rm{H}}{{\bf{A}}_{k,k}}{{\bf{w}}_k}$ in C4 with the right side of (\ref{ehdliu}), Problem ${\cal P}_5$ is transformed to the following optimization problem
\begin{subequations}\label{jaleuirot}
\begin{align}
{\cal P}_6:\ \mathop {\min }\limits_{{\bf{w}} }\quad
& \sum\nolimits_{k \in {\cal U}} {\left\| {{{\bf{w}}_k}} \right\|_2^2}  \\
\qquad\ \textrm{s.t.}\qquad\!\!\!\!
&{\rm{C2}}, {\rm{C7}}, \\
&{\rm{C8:}}\ 2{\rm{Re}}\left( {{\bf{w}}_k^{\rm{H}}(t){{\bf{A}}_{k,k}}{{\bf{w}}_k}} \right) - {\zeta _k}(t) \ge \nonumber\\
&\qquad\quad {\eta _{k,\min }}\left( {{\bf{w}}_k^{\rm{H}}{{\bf{E}}_{k,k}}{{\bf{w}}_k} + \sum\nolimits_{l \ne k,l \in {\cal U}} {{\bf{w}}_l^{\rm{H}}{{\bf{A}}_{l,k}}{{\bf{w}}_l} + \sigma _k^2} } \right),\forall k \in {\cal U},
\end{align}
\end{subequations}
where ${\zeta _k}(t) = {\bf{w}}_k^{\rm{H}}(t){{\bf{A}}_{k,k}}{{\bf{w}}_k}(t)$. Now, Problem ${\cal P}_6$ is a convex optimization problem. Additionally, in Appendix \ref{fjreeo}, we prove that the Slater's condition \cite{boyd2004convex} of Problem  ${\cal P}_6$ is satisfied. Hence, the duality gap between Problem  ${\cal P}_6$ and its dual problem is zero. As a result, the original Problem ${\cal P}_6$ can be solved by solving its dual problem instead. In the following, we derive the  structure of the optimal beamforming vector by applying the Lagrange dual decomposition method.

Let us represent ${{\cal I}_k}$ as ${{\cal I}_k} = \{ s_1^k, \cdots ,s_{|{{\cal I}_k}|}^k\}$. We first introduce the following block-diagonal matrices
\begin{equation}\label{blocmat}
  {{\bf{B}}_{i,k}}{\rm{ = diag}}\left\{ {\overbrace {{{\bf{0}}_{1 \times M}}}^{s_1^k}, \cdots ,\overbrace {{{\bf{1}}_{1 \times M}}}^{s_m^k},\overbrace {{{\bf{0}}_{1 \times M}}}^{s_{m + 1}^k}, \cdots ,\overbrace {{{\bf{0}}_{1 \times M}}}^{s_{\left| {{{\cal I}_{ k}}} \right|}^k}} \right\},\ {\rm{if }}\ s_m^k = i,\forall i \in {\cal I},k \in {\cal U}.
\end{equation}
Then,  Constraints C2 and C7 can be rewritten as
\begin{eqnarray}
&&{\rm{C9:}}\sum\nolimits_{k \in {{\cal U}_i}} {{\bf{w}}_k^{\rm{H}}{{\bf{B}}_{i,k}}{{\bf{w}}_k}}  \le {P_{i,\max }},\forall i\in\cal I\\
&&{\rm{C10:}}\sum\nolimits_{k \in {{\cal U}_i}} {{\tau _{i,k}}(t){\bf{w}}_k^{\rm{H}}{{\bf{B}}_{i,k}}{{\bf{w}}_k}}  \le {{\tilde C}_i}(t),\forall i \in {\cal I}.
\end{eqnarray}
After some further manipulations, the Lagrangian function of Problem ${\cal P}_6$ can be written as
\begin{eqnarray}
&&{\cal L}\left( {{\bf{w}},{\bm{\lambda}} ,{\bm{\mu}} ,{\bm{\nu}} } \right)\nonumber\\
 &=& \sum\limits_{k \in {\cal U}} {{\bf{w}}_k^{\rm{H}}{{\bf{J}}_k}(t){{\bf{w}}_k}}  - \sum\limits_{k \in {\cal U}} {{\upsilon _k}\left[ {{\bf{w}}_k^{\rm{H}}(t){{\bf{A}}_{k,k}}{{\bf{w}}_k} + {\bf{w}}_k^{\rm{H}}{{\bf{A}}_{k,k}}{{\bf{w}}_k}(t)} \right]}  - \sum\limits_{i \in {\cal I}} {{\lambda _i}{P_{i,\max }}}  - \sum\limits_{i \in {\cal I}} {{\mu _i}{{\tilde C}_i}(t)} \nonumber \\
 &&\   + \sum\limits_{k \in {\cal U}} {{\upsilon _k}\left[ {{\eta _{k,\min }}\sigma _k^2 + {\zeta _k}(t)} \right]} ,\nonumber
\end{eqnarray}
where ${\bm{\lambda}} ,{\bm{\mu}} ,{\bm{\nu}}$ are the collections of non-negative Lagrangian multipliers associated with Constraint C9, C10 and C8, respectively, the matrix ${{{\bf{J}}_k}}(t)$ above is given by
\begin{equation}\label{hfiuere}
{{\bf{J}}_k}(t) = {\bf{I}} + \sum\nolimits_{i \in {{\cal I}_k}} {\left( {{\lambda _i} + {\mu _i}{\tau _{i,k}}(t)} \right){{\bf{B}}_{i,k}}}  + {\upsilon _k}{\eta _{k,\min }}{{\bf{E}}_{k,k}} + \sum\nolimits_{l \ne k,l \in {\cal U}} {{\eta _{l,\min }}{{\bf{A}}_{k,l}}}.
\end{equation}
Then, the dual function is given by
\begin{eqnarray}
&&g\left( {{\bm{\lambda}} ,{\bm{\mu}} ,{\bm{\nu}}} \right)\\
 &=& \mathop {\min }\limits_{{\bf{w}}} {\cal L}\left( {{\bf{w}},{\bm{\lambda}} ,{\bm{\mu}} ,{\bm{\nu}} } \right)\\
&=& \mathop {\min }\limits_{{\bf{w}}}  \sum\nolimits_{k \in {\cal U}} {{\bf{w}}_k^{\rm{H}}{{\bf{J}}_k}(t){{\bf{w}}_k}}  - \sum\nolimits_{k \in {\cal U}} {{\upsilon _k}\left[ {{\bf{w}}_k^{\rm{H}}(t){{\bf{A}}_{k,k}}{{\bf{w}}_k} + {\bf{w}}_k^{\rm{H}}{{\bf{A}}_{k,k}}{{\bf{w}}_k}(t)} \right]} \nonumber\\
&&\qquad - \sum\nolimits_{i \in {\cal I}} {{\lambda _i}{P_{i,\max }}}  - \sum\nolimits_{i \in {\cal I}} {{\mu _i}{{\tilde C}_i}(t)} + \sum\nolimits_{k \in {\cal U}} {{\upsilon _k}\left[ {{\eta _{k,\min }}\sigma _k^2 + {\zeta _k}(t)} \right]}.\label{rjfoerhg}
\end{eqnarray}
Note that ${{\bf{J}}_k}(t)$ is a positive definite matrix. Hence, Problem (\ref{rjfoerhg}) is a strictly convex problem and its unique solution can be obtained  from its first-order optimality condition as:
 \begin{equation}\label{tiuotri}
   {{\bf{w}}_k} = {\upsilon _k}{\bf{J}}_k^{ - 1}(t){{\bf{A}}_{k,k}}{{\bf{w}}_k}(t).
 \end{equation}
By substituting the optimal solution of ${{\bf{w}}_k}$ in (\ref{tiuotri}) into (\ref{rjfoerhg}), the dual function becomes
\begin{eqnarray}
g({\bm{\lambda}} ,{\bm{\mu}} ,{\bm{\nu}}) &=&   - \sum\nolimits_{k \in {\cal U}} {\upsilon _k^2{\bf{w}}_k^{\rm{H}}(t){{\bf{A}}_{k,k}}{\bf{J}}_k^{ - 1}(t){{\bf{A}}_{k,k}}{{\bf{w}}_k}(t)}  - \sum\nolimits_{i \in {\cal I}} {{\lambda _i}{P_{i,\max }}}  - \sum\nolimits_{i \in {\cal I}} {{\mu _i}{{\tilde C}_i}(t)}  \nonumber\\
 &&  + \sum\nolimits_{k \in {\cal U}} {{\upsilon _k}\left[ {{\eta _{k,\min }}\sigma _k^2 + {\zeta _k}(t)} \right]}. \label{dualfunc}
\end{eqnarray}
Then, the dual of Problem ${\cal P}_6$ is given by
\begin{equation}\label{dualproblem}
  \mathop {\max }\limits_{\left\{ {{\lambda _i} \ge 0,{\mu _i} \ge 0,{\nu _k} \ge 0,\forall k,i} \right\}} g({\bm{\lambda}} ,{\bm{\mu}} ,{\bm{\nu}} ).
\end{equation}
The classic gradient descent methods such as  the subgradient or ellipsoid methods \cite{boyd2004convex} can be employed to solve the dual problem (\ref{dualproblem}) to update the Lagrangian multipliers.

\subsection{Low-complexity Algorithm}\label{fjjfrijojoy}
Combining Subsection-\ref{sonemo} and Subsection \ref{freito}, we conceive an iterative algorithm to solve Problem ${\cal P}_3$ based on the first order Taylor approximation (FOTA) method in Algorithm \ref{algorithmiterFOTA}. It is readily seen that the optimal solution obtained at the $t$th iteration is also feasible for Problem ${\cal P}_3$ at the $(t+1)$th iteration, since the indicator function is smaller than one and it is approximated as the right hand side of (\ref{fucntions}). This implies that Algorithm \ref{algorithmiterFOTA} generates a non-increasing sequence of objective function values and finally converges to the Karush-Kuhn-Tucker solution of
Problem ${\cal P}_4$, as proved in \cite{cunhua2015wcl}. Note that the optimal beamforming solution obtained by Algorithm \ref{algorithmiterFOTA} is guaranteed to be rank one.

In Algorithm \ref{algorithmiterFOTA}, it is necessary to find the initial feasible set of beamforming vectors ${\bf{w}}{(0)}$. In Section \ref{userselec}, we provide the UE selection algorithm to find the maximum number of admitted UEs. The corresponding obtained beamforming vectors can be set as the initial point of Algorithm \ref{algorithmiterFOTA}. The reason is that the constraints of Problem ${\cal P}_3$ and Problem ${\cal P}_7$ are the same.

\begin{algorithm}
\caption{FOTA-based Algorithm to Solve Problem ${\cal P}_3$}\label{algorithmiterFOTA}
\begin{algorithmic}[1]
\STATE Initialize   iteration number $t=1$,  error tolerance $\delta $, small constant $\theta$, feasible ${\bf{w}}{(0)}$, calculate ${\tau _{i,k}}(0)$, ${{\tilde C}_i}(0)$ and ${\zeta _k}(0)$,  calculate the objective value of Problem ${\cal P}_6$, denoted as ${\rm{Obj}}(0)$.
 \STATE Solve Problem ${\cal P}_6$ by using the Lagrange dual decomposition method to obtain $\left\{ {{{\bf{w}}_k}(t),\forall k} \right\}$ with ${\tau _{i,k}}(t-1)$, ${{\tilde C}_i}(t-1)$ and $\zeta _k(t-1)$;
 \STATE With $\left\{ {{{\bf{w}}_k}(t),\forall k} \right\}$, update ${\tau _{i,k}}(t)$, ${{\tilde C}_i}(t)$  and $\zeta _k(t)$;
 \STATE If  ${{\left| {{\rm{Obj(}}t - 1{\rm{) - Obj(}}t{\rm{)}}} \right|} \mathord{\left/
 {\vphantom {{\left| {{\rm{Obj(}}t - 1{\rm{) - Obj(}}t{\rm{)}}} \right|} {{\rm{Obj(}}t{\rm{)}}}}} \right.
 \kern-\nulldelimiterspace} {{\rm{Obj(}}t{\rm{)}}}}< \delta  $, terminate.  Otherwise, set $t \leftarrow t + 1$, go to step 2.
\end{algorithmic}
\end{algorithm}
\vspace{-0.5cm}\subsection{Complexity Analysis}\label{ehafuir}
In this subsection, we analyze the computational complexity of Algorithm \ref{algorithmiterFOTA}. For notational simplicity, we assume that  candidate set size for each UE is equal to $L$, $|{\cal I}_k|=L, \forall k \in \cal U$. Note that in general $L$ is much smaller than the total number of RRHs $I$.

For Algorithm \ref{algorithmiterFOTA}, the main complexity lies in solving Problem ${\cal P}_6$ by using the Lagrange dual decomposition method. In each iteration of the Lagrange dual decomposition method, the complexity is dominated by calculating $ {{\bf{w}}_k}$ in (\ref{tiuotri}). Note that the complexity of calculating $ {{\bf{w}}_k}$ mainly lies in the calculation of ${\bf{J}}_k^{ - 1}(t)$. According to \cite{boyd2004convex}, for a complex matrix ${\bf{A}}\in {\mathbb{C}}^{N \times N}$, the complexity of calculating ${{\bf{A}}^{ - 1}}$ is on the order of $O(N^3)$. Hence, the complexity of calculating $ {{\bf{w}}_k}$ is on the order of $O(M^3L^3)$. Since there are a total of $K$ UEs, the total complexity of the Lagrange dual decomposition method in each iteration is on the order of $O(KM^3L^3)$. Since there are a total of $(2I+K)$ dual variables, the total number of iterations required by the ellipsoid methods is upper-bounded by  $O[(2I+K)^2]$ \cite{ben2001lectures}. Hence, the total complexity of the Lagrange dual decomposition method is given by $O[(2I+K)^2KM^3L^3]$. Let us denote $t_{\rm{avg}}$ as the average number of iterations required for Algorithm \ref{algorithmiterFOTA} to converge, then the total complexity of Algorithm \ref{algorithmiterFOTA} imposed by solving Problem ${\cal P}_1$ is expressed as $T_{{\cal P}_1}=O[t_{\rm{avg}}(2I+K)^2KM^3L^3]$. Simulation results show that Algorithm \ref{algorithmiterFOTA} converges  fast, typically 10 iterations are sufficient for the algorithm to converge.

\section{Low-complexity UE Selection Algorithms}\label{userselec}

In this section we solve the UE selection Problem ${\cal P}_1$. By substituting $r_k=R_{k,\rm{min}},\forall k$ into the fronthaul capacity constraint C3, we obtain an alternative optimization problem to Problem ${\cal P}_1$, which is expressed as follows:
\begin{equation}\label{usefreo}
\begin{array}{l}
 {\cal P}_7:\ \mathop {\max }\limits_{{\bf{w}}, {\cal U} \subseteq {\overline {\cal U}} } \quad \left| \cal U \right|\\
\qquad\ {\rm{s}}.{\rm{t}}.\qquad {\kern 1pt} {\rm{C2}},{\rm{C4}},{\rm{C5}},
\end{array}
\end{equation}
where C4 and C5 are given in Problem ${\cal P}_3$. Although Problem ${\cal P}_7$ is not the same as the original UE selection Problem ${\cal P}_1$, Problem ${\cal P}_7$ is equivalent to Problem ${\cal P}_1$ in the sense that both problems yield the same optimal set of selected UEs, the proof of which can be found in Appendix \ref{hfdewafreh}. It should be noted that the optimal beamforming vectors obtained from solving Problem ${\cal P}_7$ may not be feasible for Problem ${\cal P}_1$. However, the aim of solving Problem ${\cal P}_7$ is twofold. Firstly, one can find the optimal set of selected UEs. Second, one can provide the initial feasible point for solving Problem ${\cal P}_3$ in Stage II since both problems have the same set of constraints.

Inspired by the UE selection method of \cite{Matskani2008}, we construct an alternative  to Problem ${\cal P}_7$ by introducing a set of auxiliary variables $\{\varphi_k\}_{k\in {\bar {\cal U}}} $:
\begin{subequations}\label{jalejoihjyoi}
\begin{align}
{\cal P}_8:\ \mathop {\min }\limits_{\{\varphi_k\geq 0\}_{k\in {\overline {\cal U}}}, \bf{w}}\quad
& \sum\nolimits_{k \in \overline {\cal U}} {\varphi_k}  \\
\qquad\ \textrm{s.t.}\qquad\!\!\!\!
&{\rm{C2}}, {\rm{C5}}, \\
&{\rm{C11:}}\ {{\bf{w}}_k^{\rm{H}}{{\bf{A}}_{k,k}}{{\bf{w}}_k}}+\varphi_k\geq \nonumber\\
&\qquad\  {\eta _{k,\min }}\left( {{\bf{w}}_k^{\rm{H}}{{\bf{E}}_{k,k}}{{\bf{w}}_k} + \sum\nolimits_{l \ne k,l \in {\cal U}} {{\bf{w}}_l^{\rm{H}}{{\bf{A}}_{l,k}}{{\bf{w}}_l} + \sigma _k^2} } \right),\forall k\in {\overline {\cal U}}.
\end{align}
\end{subequations}
Let us denote the solution  of $\{\varphi_k\}_{k\in {\overline{\cal U}}}$ by ${\{ \varphi _k^\star\} _{k \in \overline{\cal U} }}$. It is readily seen that Problem ${\cal P}_8$ is always feasible. If the optimal solutions of $\{\varphi_k^\star\}_{ k\in \overline{\cal U}}$ are all equal to zero, then all UEs can be admitted to the network. Otherwise, some UEs should be removed from the system and we reschedule them for the next opportunity. Intuitively, the UE having a largest value of $\varphi_k^\star$ has a higher probability to be removed since it has the largest discrepancy from its rate target. Problem ${\cal P}_8$ can be solved similarly  to Problem ${\cal P}_3$ of the above section, hence the details of which are omitted.

There are two low-complexity UE deletion methods. One is the successive UE deletion method that is provided in \cite{Matskani2008,pan2017twc}. The main idea is to remove the UE having the largest $\varphi_k^\star$  each time, until all the remaining optimal values of $\varphi_k^\star$ become equal to zero. The complexity of this algorithm increases linearly with the number of UEs, hence it is on the order of $O(K)$. This algorithm is suitable for medium-sized networks. The other technique is the bisection based search method  proposed in \cite{pan2017joint}. The main idea is to sort ${\{ \varphi _k^\star\} _{k \in  \overline{\cal U} }}$ in descending order $\varphi _{{\pi _1}}^\star \ge  \cdots  \ge \varphi _{{\pi _K}}^\star$. Then, one should find a minimum ${L_0}$ for ensuring that all the UEs in $ {\cal U}  = \left\{ {{\pi _{{L_0} + 1}}, \cdots ,{\pi _K}} \right\}$ can be supported with ${L_0} = 1, \cdots ,K - 1$. The bisection search method is used to iteratively find the optimal ${L_0}$ by updating its upper-bound and lower-bound. The complexity of the bisection based method is on the order of $\left\lceil {{{\log }_2}(1 + K)} \right\rceil $, which is suitable for very dense networks supporting a large number of UEs. The details of these two algorithms are not shown here for simplicity.

It should be emphasized that when using the iterative algorithm in the above section to solve Problem ${\cal P}_8$, the iterative procedure will terminate once the intermediate solutions of $\{\varphi_k\}_{ k\in {\cal U}}$ are all equal to zero. Hence, the data rates of some UEs with the obtained beamforming solution are strictly larger than their minimum rate requirements.

\section{Simulation Results}\label{simlresult}
In this section, we provide simulation results to evaluate the performance of the proposed robust algorithms. Two types of UD-CRAN networks are considered: a small UD-CRAN deployed in a square area of [400 m $\times$ 400 m] and a larger one of [700 m $\times$ 700 m].  Both the UEs and RRHs are uniformly distributed in these areas. For the small one, the numbers of RRHs and UEs are set to $I=14$ and $K=8$ with the densities of 87.5 RRHs/km$^2$ and 50 UEs/km$^2$, respectively. For the large one, the numbers are set to $I=42$ and $K=24$ with the densities of 85.7 RRHs/km$^2$ and 49 UEs/km$^2$, respectively. These two scenarios comply with the ultra-dense networks in the fifth-generation (5G) wireless system \cite{xiaohuge2016}, where the density of BSs will be up to 40-50 BSs/km$^2$. The channels are generated according to the LTE specifications \cite{access2010further}, which are composed of three elements: 1) the large-scale path loss given by $PL = 148.1 + 37.6{\log _{10}}d\ ({\rm{dB}})$, where $d$ is the distance between a RRH and a UE in km; 2) the log-normal shadowing fading having a zero mean and 8 dB standard deviation;
3) small-scale Rayleigh fading with zero mean and unit variance. For ease of exposition, all UEs are assumed to have the same rate constraints of ${R_{\min }} = {R_{k,\min }}, \forall k$, and all RRHs have the same power constraints of $P_{\rm{max}}=P_{i,\rm{max}}, \forall i$. Furthermore, the fronthaul capacity constraints are assumed to be the same for all RRHs, i.e., ${C_{\max }} = {C_{i,\max }},\forall i$, and we consider the normalized fronthaul capacity constraints (with respect to each UE's rate traget), i.e., ${{\tilde C}_{\max }} = {{{C_{\max }}} \mathord{\left/
 {\vphantom {{{C_{\max }}} {{R_{\min }}}}} \right.
 \kern-\nulldelimiterspace} {{R_{\min }}}}$.  Note that ${{\tilde C}_{\max }}$ can be interpreted as the maximum number of UEs that can be supported by each fronthaul link. For simplicity, each UE is assumed to choose its nearest $L$ RRHs as its serving candidate set, i.e., $\left| {{{\cal I}_k}} \right| = L,\forall k$. The maximum pilot reuse times for small and large UD-CRANs are $n_{\rm{max}}=2$  and $n_{\rm{max}}=3$, respectively.
The total number of time slots in each time frame is $T=200$, the numbers of CDI and PA quantization bits for each RRH are set as  $B^{{\rm{CDI}}}=4$ and $B^{{\rm{PA}}}=2$, respectively. Unless otherwise stated, the simulation parameters are given in Tabel \ref{tab3} and the following results are obtained by averaging over 100 channel generations.
\begin{table}[!t]
\renewcommand{\arraystretch}{1.1}
\caption{Main simulation parameters}\vspace{-0.5cm}
\label{tab3}
\centering
\begin{tabular}{c|c|c|c}
\hline\hline
\textbf{Parameters}  & \textbf{Value}&\textbf{Parameters}  & \textbf{Value}  \\
 \hline
 Number of antennas  $M$ & 2 & System bandwidth $B$ &  20 ${\rm{MHz}}$   \\
 \hline
  Noise power density& -174 dBm/Hz \cite{access2010further} & Error tolerance $\delta$ &  $10^{-5}$ \\
\hline
Small constant $\theta$& $10^{-5}$ & Pilot power $p_t$ & 200 mW\\
\hline
Maximum transmit power $P_{\rm{max}}$ &  100 mW & Rate target ${R_{\min }}$ & $3\ \rm{bit/s/Hz}$\\
\hline
Candidate size $L$ & 3 & Normalized fronthaul limits ${{\tilde C}_{\max }}$& 3\\
\hline\hline
\end{tabular}\vspace{-0.5cm}
\end{table}
\subsection{Smaller UD-CRAN}

\begin{figure}
\begin{minipage}[t]{0.475\linewidth}
\centering
\includegraphics[width=2.6in]{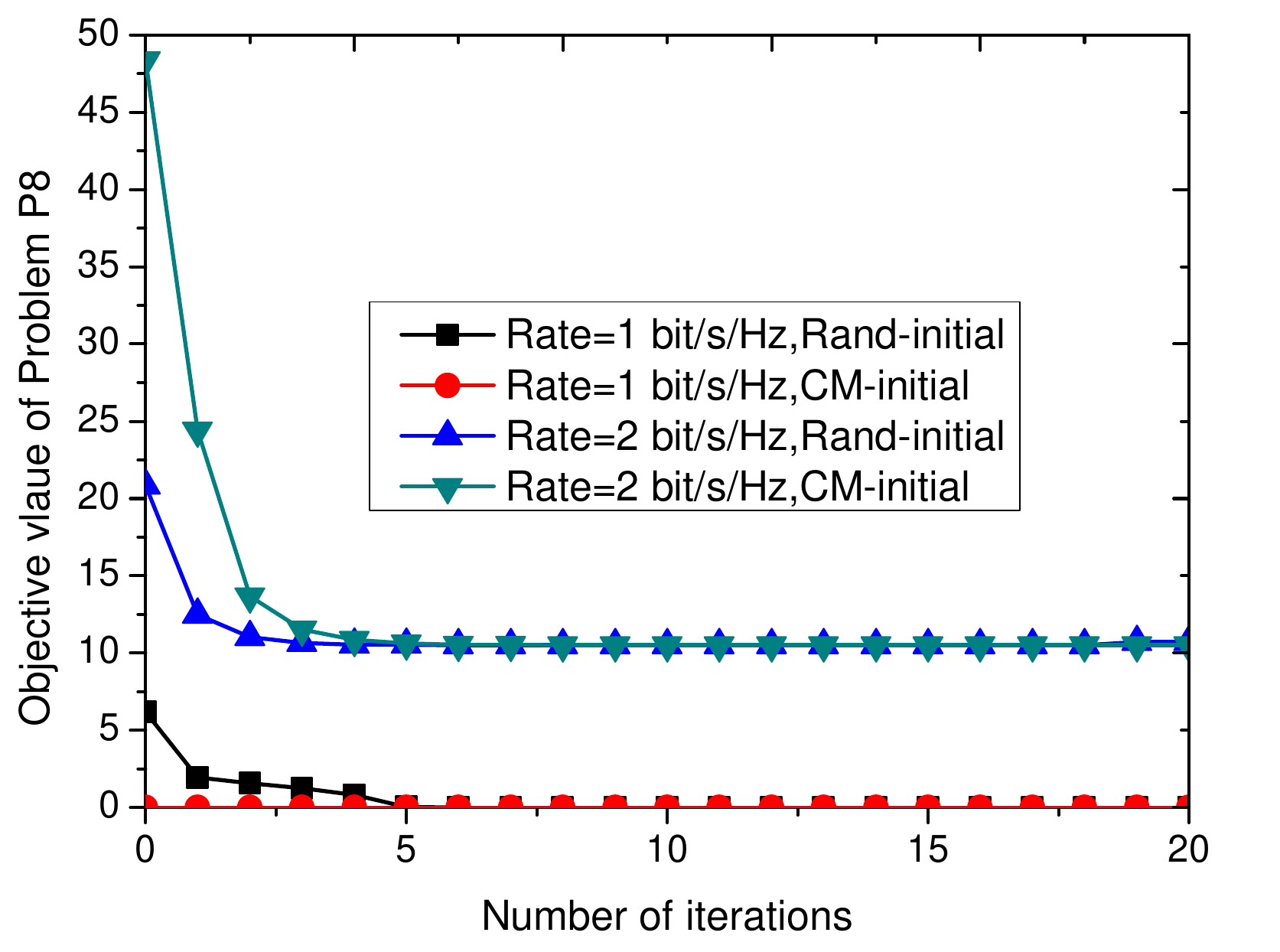}\vspace{-0.4cm}
\caption{Convergence behaviour of the FOTA-based Algorithm to solve Problem ${\cal P}_8$ under different initial points.}\vspace{-0.8cm}
\label{coninitial}
\end{minipage}%
\hfill
\begin{minipage}[t]{0.475\linewidth}
\centering
\includegraphics[width=2.6in]{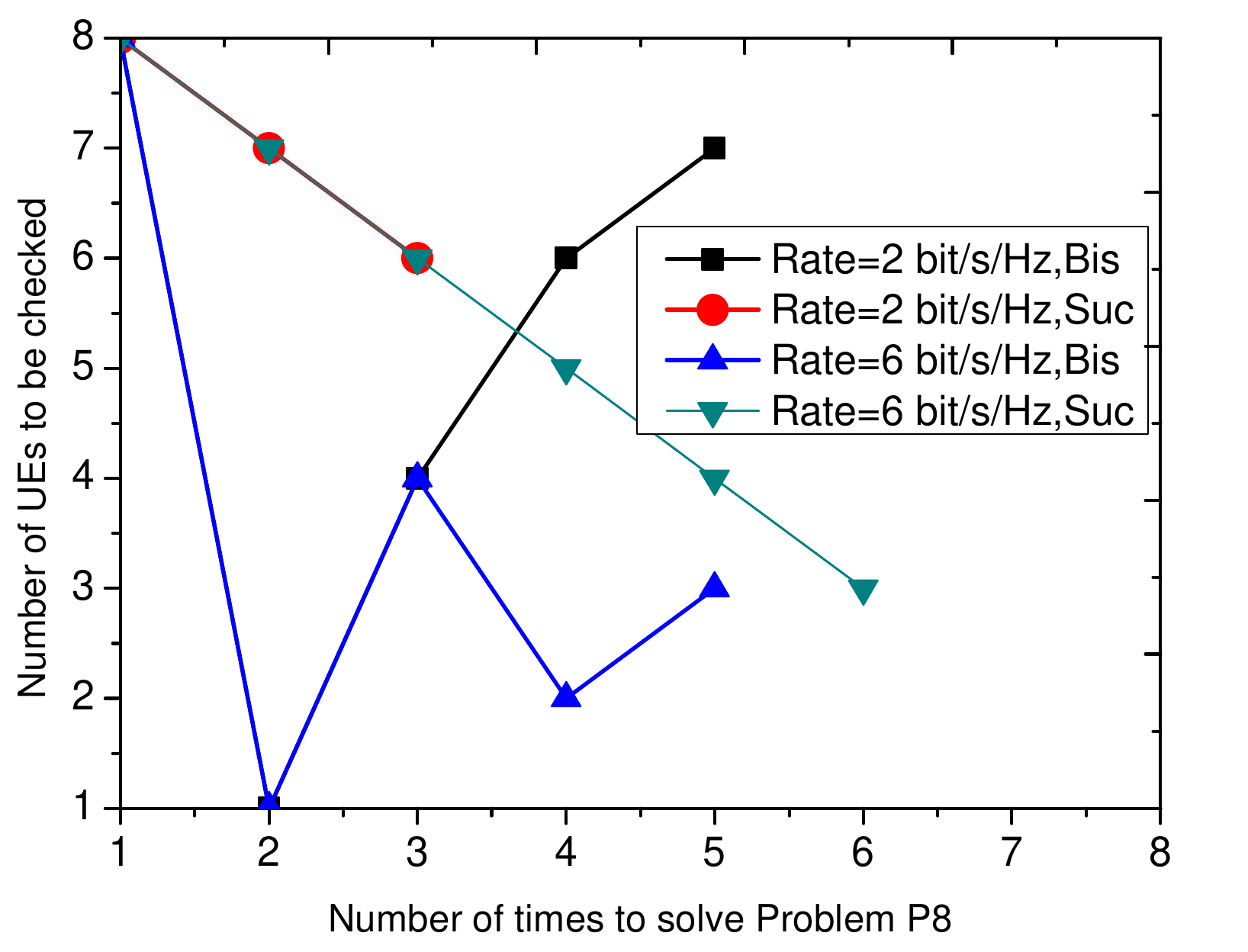}\vspace{-0.4cm}
\caption{Convergence behaviour of the proposed two UE selection algorithms.}\vspace{-0.8cm}
\label{contwouesel}
\end{minipage}%
\hfill
\end{figure}

\begin{figure}
\begin{minipage}[t]{0.475\linewidth}
\centering
\includegraphics[width=2.6in]{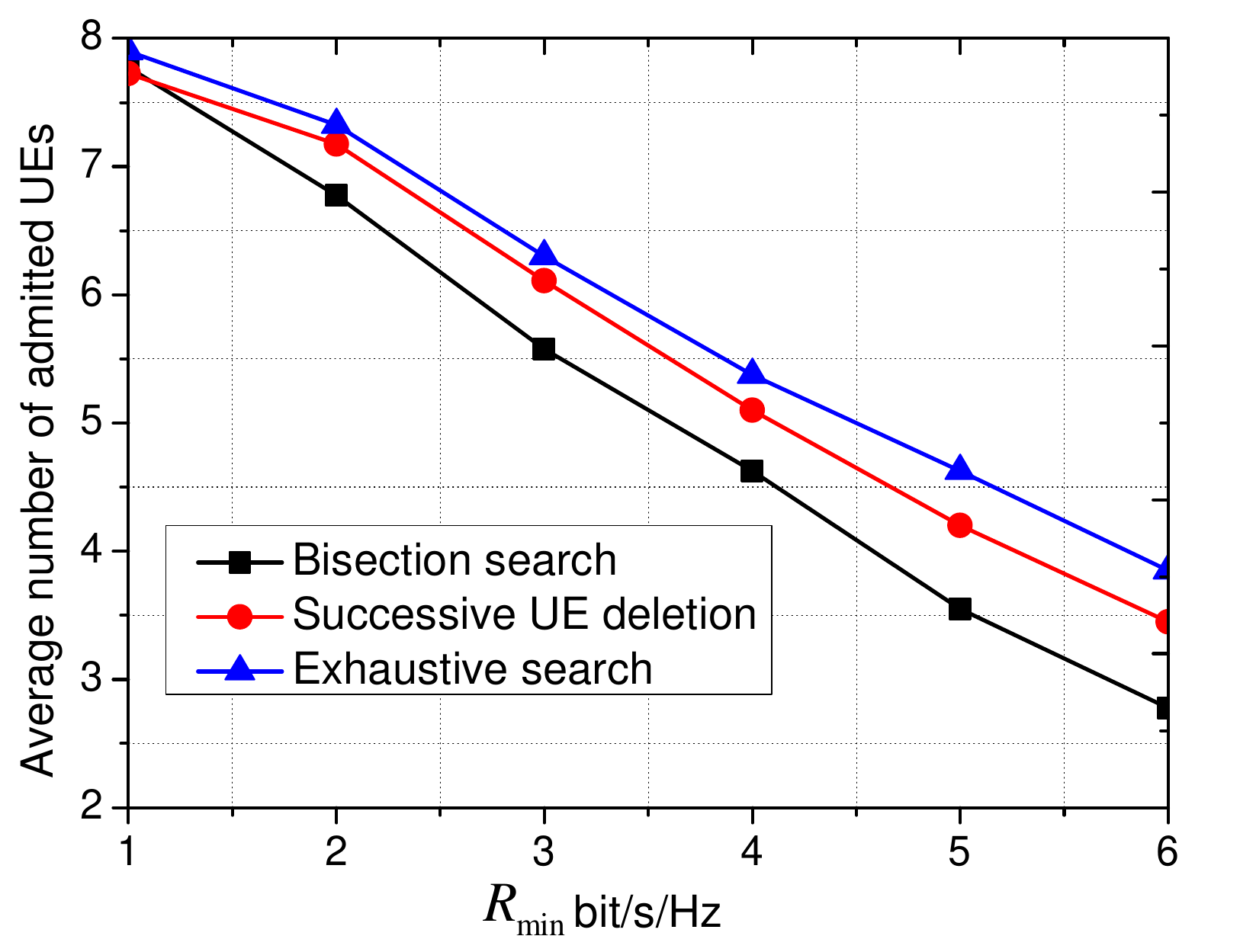}\vspace{-0.4cm}
\caption{Number of UEs admitted  by various algorithms versus different rate targets.}\vspace{-0.9cm}
\label{fig3}
\end{minipage}%
\hfill
\begin{minipage}[t]{0.475\linewidth}
\centering
\includegraphics[width=2.6in]{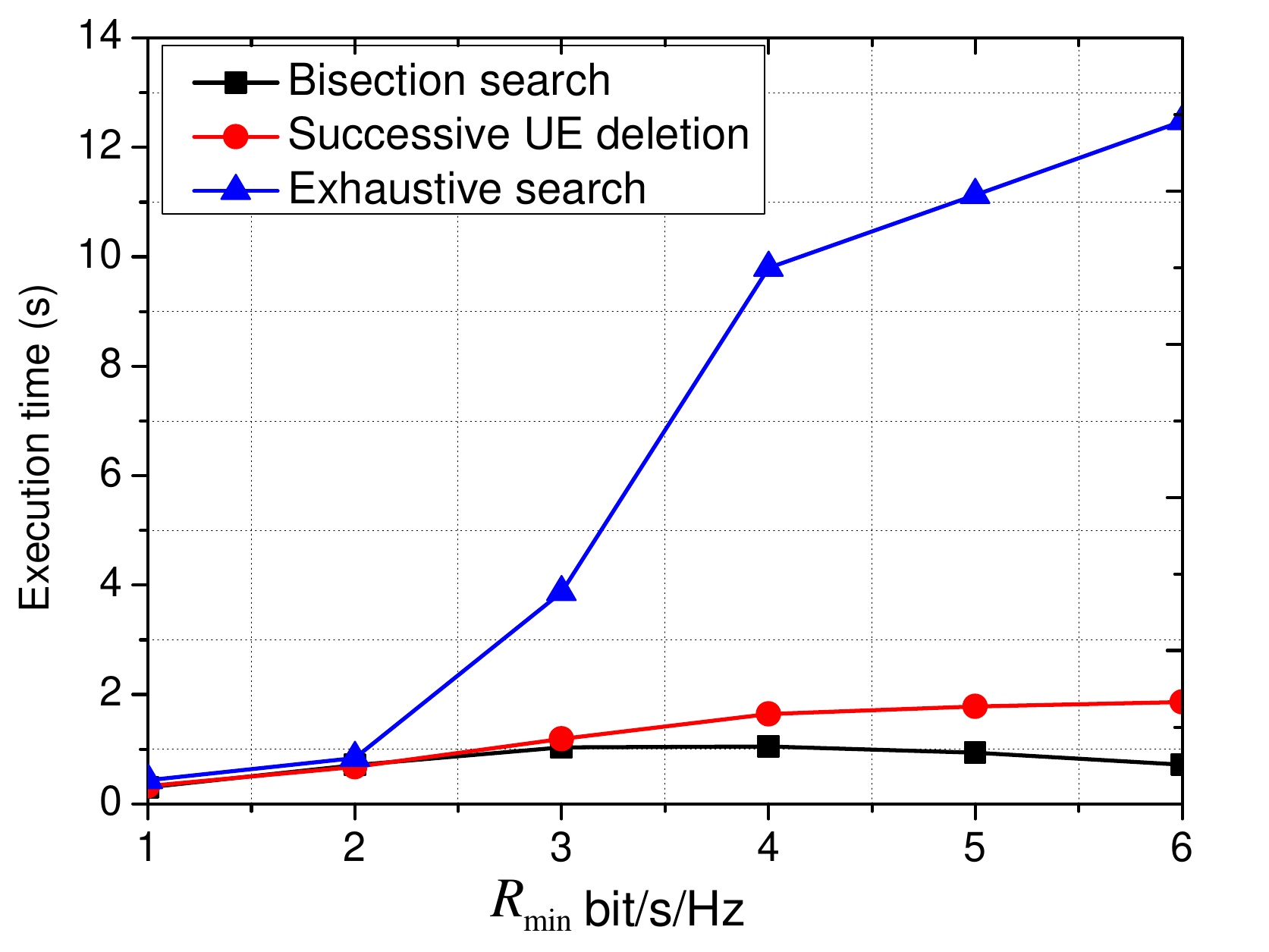}\vspace{-0.4cm}
\caption{Execution time for various UE selection algorithms.}\vspace{-0.9cm}
\label{fig4}
\end{minipage}%
\hfill
\end{figure}

\begin{figure}
\begin{minipage}[t]{0.475\linewidth}
\centering
\includegraphics[width=2.6in]{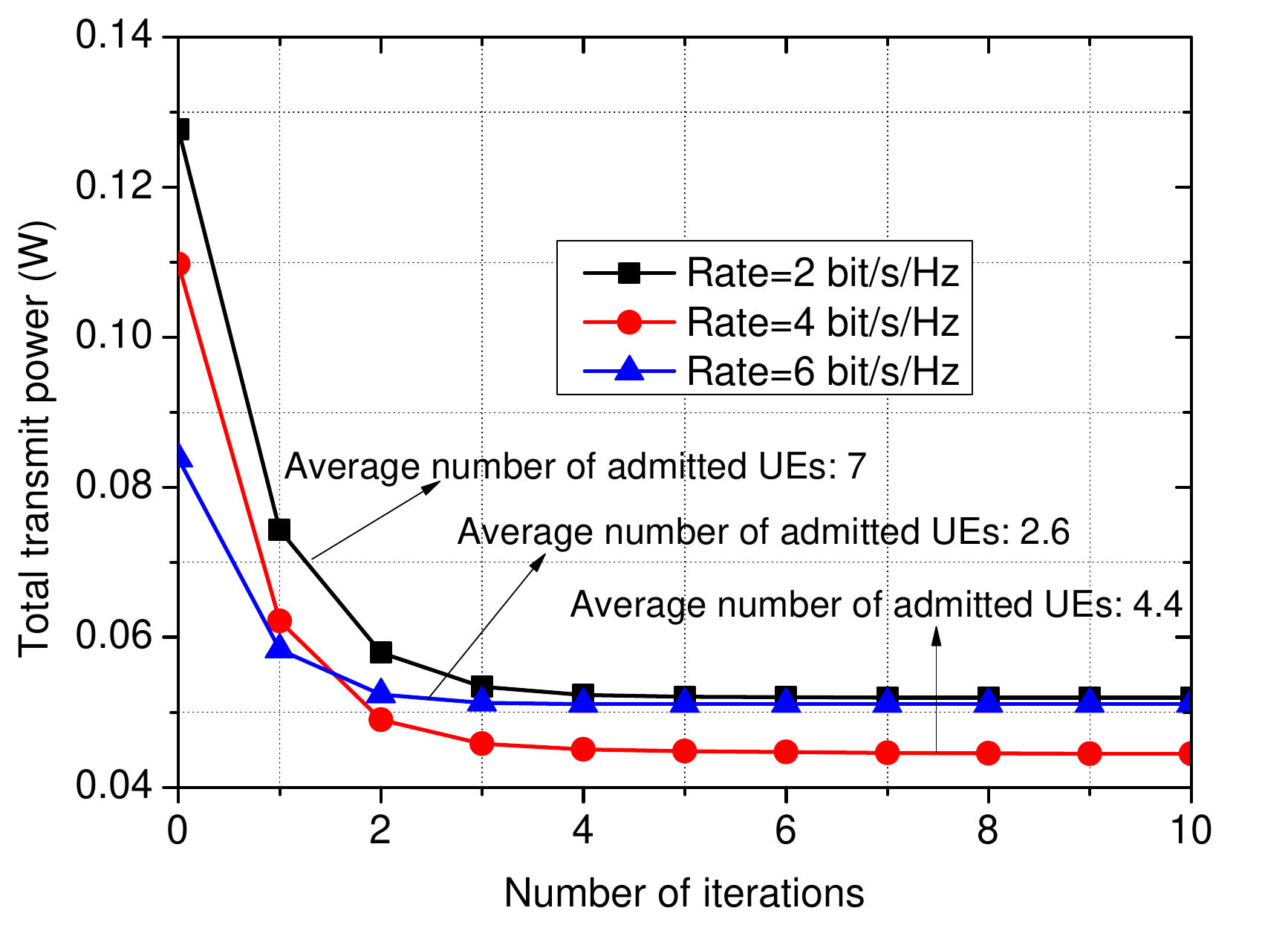}\vspace{-0.4cm}
\caption{Convergence behaviour of the FOTA-based Algorithm to solve Problem ${\cal P}_3$ under different rate targets.}\vspace{-0.6cm}
\label{figcon}
\end{minipage}%
\hfill
\begin{minipage}[t]{0.475\linewidth}
\centering
\includegraphics[width=2.6in]{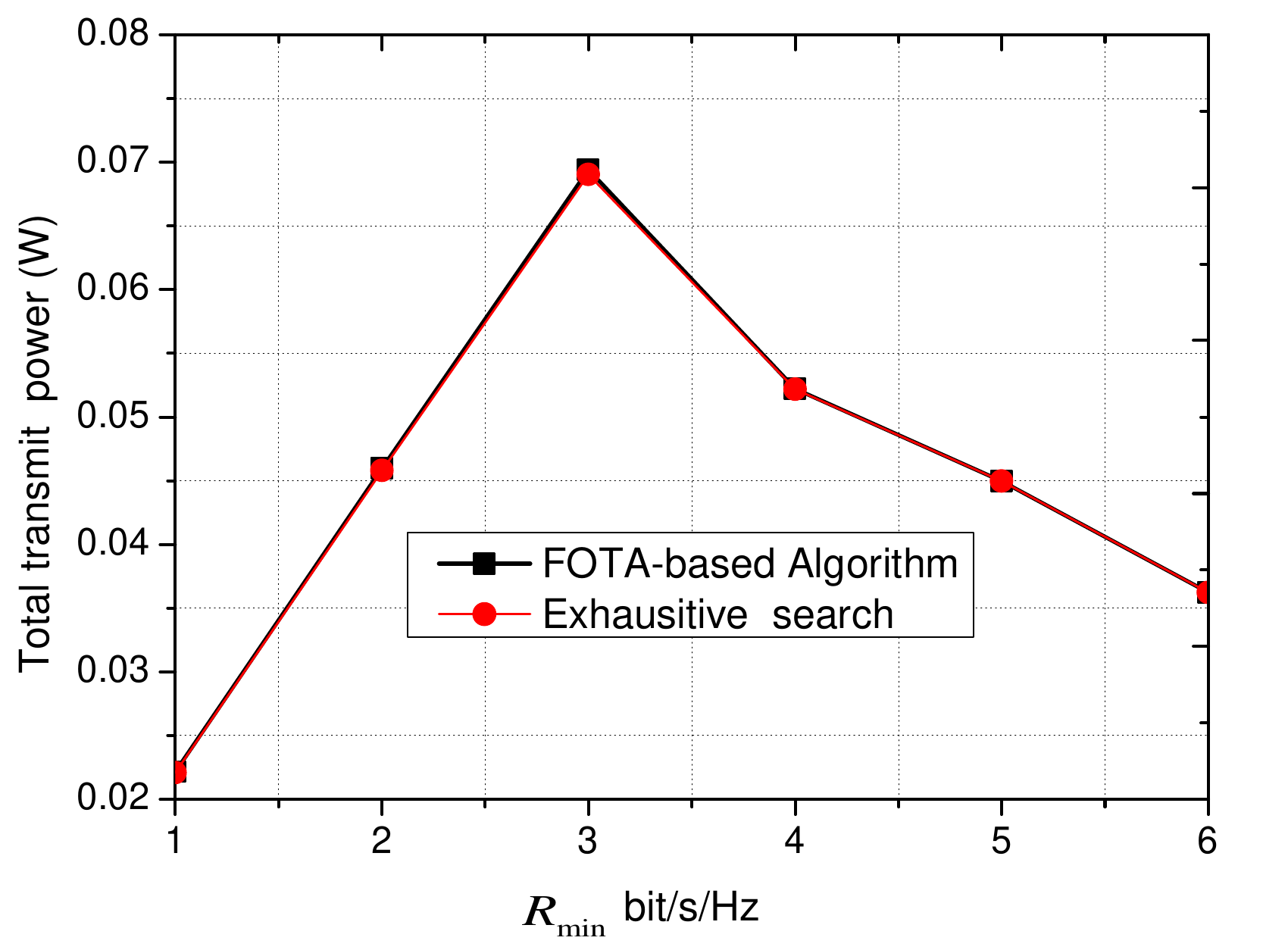}\vspace{-0.4cm}
\caption{Total transmit power versus the rate targets for the proposed algorithm and the exhaustive search method.}\vspace{-0.6cm}
\label{powvsrate}
\end{minipage}%
\hfill
\end{figure}

\begin{figure}
\begin{minipage}[t]{0.475\linewidth}
\centering
\includegraphics[width=2.6in]{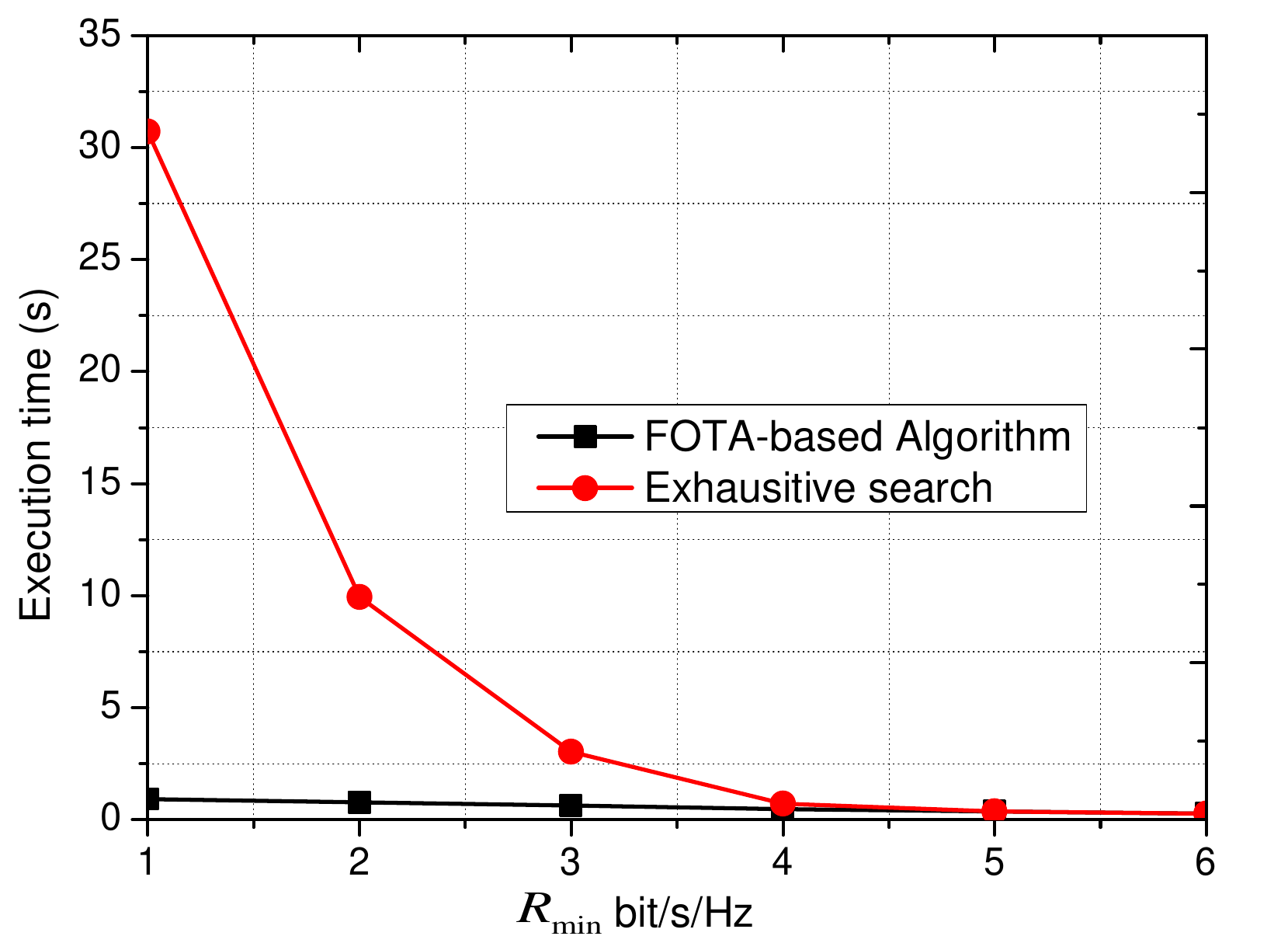}\vspace{-0.4cm}
\caption{Execution time for the proposed FOTA-based Algorithm and the exhaustive search method.}\vspace{-0.8cm}
\label{timeex}
\end{minipage}%
\hfill
\begin{minipage}[t]{0.475\linewidth}
\centering
\includegraphics[width=2.6in]{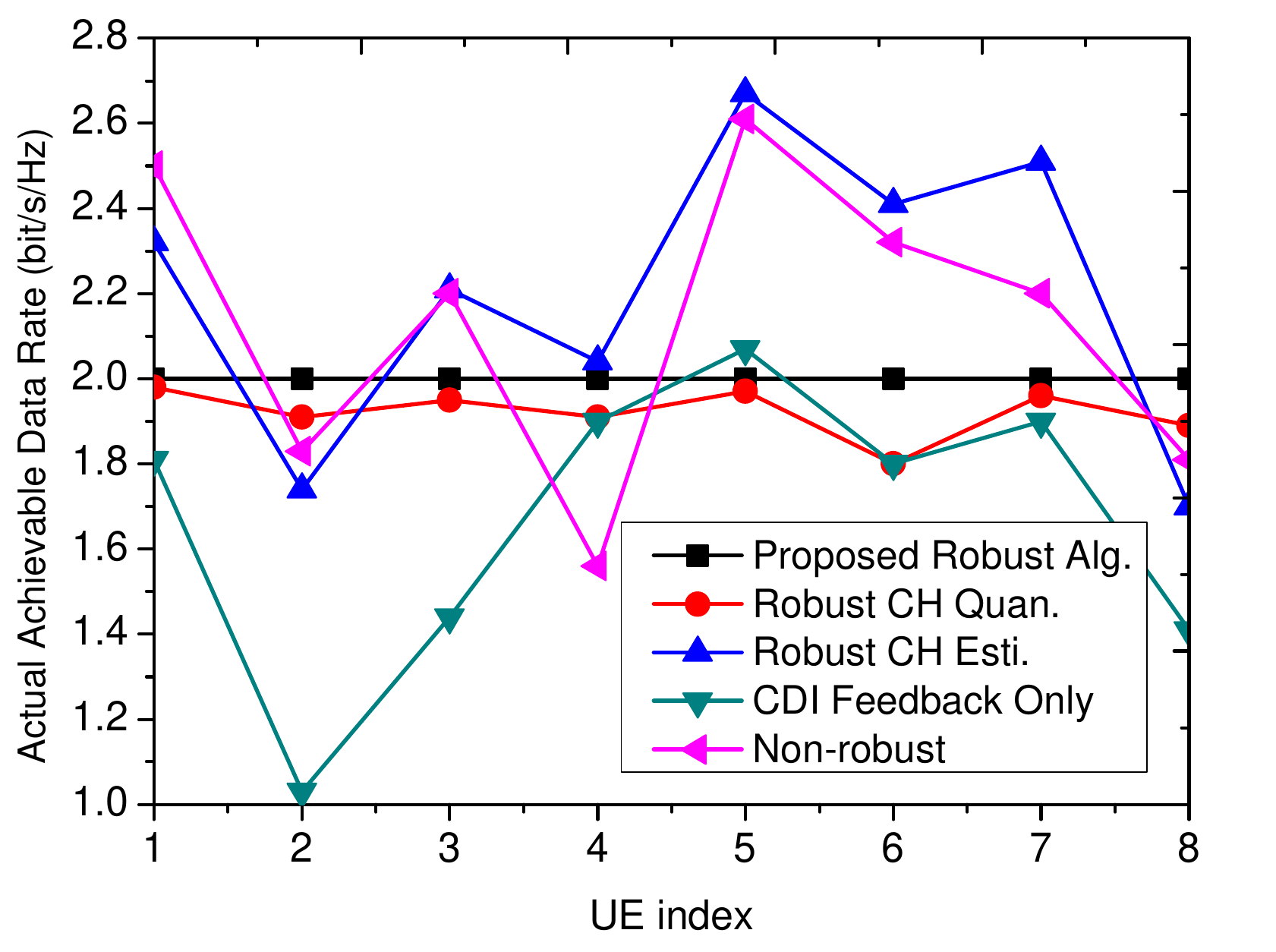}\vspace{-0.4cm}
\caption{The achievable data rate for various algorithms, where the rate target is set as $R_{\rm{min}}=2\ \rm{bit/s/Hz}$.}\vspace{-0.8cm}
\label{powvsfrfrate}
\end{minipage}%
\hfill
\end{figure}

In this section, we evaluate the performance of our algorithm in the small C-RAN network, where the simulation results in Fig.~\ref{coninitial}-Fig.~\ref{powvsfrfrate} are based on this scenario.

We first study the impact of the initial points on the convergence behaviour of the FOTA-based Algorithm to solve Problem ${\cal P}_8$. Fig.~\ref{coninitial} shows the objective value of Problem ${\cal P}_8$ versus the number of iterations for one randomly generated set of channel realizations for two cases of $R_{\rm{min}}=1\  {\rm{bit/s/Hz}}$ and $R_{\rm{min}}=2\  {\rm{bit/s/Hz}}$. Since Problem ${\cal P}_8$ is a non-convex problem, different initial points may lead to different solutions. To investigate this effect, we consider two initialization schemes: 1) Rand-initial: In this scheme, both the power allocation and beamforming direction on each beam is randomly generated; 2) CM-initial: For this scheme, the total power on each RRH is equally split among its served UEs and the beamforming direction is set to  be the same as its channel direction. It can be observed from Fig.~\ref{coninitial} that the algorithm with different initial points will have different convergence speeds, but converge to the same objective value. It is difficult to justify which initialization scheme has faster convergence speed as seen in  Fig.~\ref{coninitial}. For both schemes,  five iterations are sufficient for the algorithm to converge. When $R_{\rm{min}}=1\  {\rm{bit/s/Hz}}$, the algorithm converges to zero, which means all the UEs can be admitted. However, when $R_{\rm{min}}=2\  {\rm{bit/s/Hz}}$, the algorithm converges to a positive value, which implies that some UEs should be deleted. 

Next, we study the convergence behaviour of the proposed two UE selection algorithms. Specifically, Fig.~\ref{contwouesel} illustrates the number of UEs to be checked versus the number of times to solve Problem ${\cal P}_8$ for a randomly generated network, where the successive UE deletion method and the bisection search method are labeled as `Suc' and `Bis', respectively. It can be found from Fig.~\ref{contwouesel} that the number of UEs to be checked for the `Suc' algorithm always decreases with the number of times Problem ${\cal P}_8$  is solved, while that of the `Bis' algorithm fluctuates during the procedure. These observations are consistent with the features of these two algorithms. Interestingly, for both rate targets, the numbers of times by the `Bis' algorithm are fixed to five. However, the number of times Problem ${\cal P}_8$  is solved by the `Suc' algorithm depends on the rate targets. For the example in Fig.~\ref{contwouesel}, the `Suc' algorithm  only needs three times when $R_{\rm{min}}=2\  {\rm{bit/s/Hz}}$, while six times when $R_{\rm{min}}=6\  {\rm{bit/s/Hz}}$.

In Fig.~\ref{fig3}, we plot the number of UEs admitted by the various algorithms versus the rate targets for the smaller UD-CRAN. The exhaustive UE search algorithm (labeled as `Exhaustive search') is used as a performance benchmark, which checks all subsets of UEs and chooses the one having the largest number of admitted UEs. Note that the computational complexity of the exhaustive search is on the order of $O(2^K)$. As expected, the number of UEs admitted by all the algorithms is reduced upon increasing the UEs' data rate targets. The exhaustive search method performs better than the other two algorithms, which comes at the expense of a high computational complexity. However, its performance gain is negligible in the low rate regime. In the high data rate target regime, the performance gain of the exhaustive search method over the successive UE deletion still remains limited to $0.5$. Hence, for moderate-sized UD-CRANs, the successive UE deletion is a good option. The bisection based search method has a modest performance loss compared to the other two algorithms. Hence, the bisection based search method is more suitable for larger UD-CRAN, as a benefit of its lowest complexity.

Fig.~\ref{fig4} compares the execution time for various UE selection algorithms by using an E5-1650 CPU operating at 3.5GHz. This figure shows that for the small $R_{\rm{min}}$, all the algorithms have almost the same operation time. This phenomenon is reasonable, which can be explained as follows. In the small $R_{\rm{min}}$ regime, almost all the UEs can be admitted, hence both algorithms only need to solve Problem ${\cal P}_8$ for once. However, for the large $R_{\rm{min}}$, the exhaustive search algorithm needs significantly higher operation time than the proposed two UE selection algorithms, and the gap increases with $R_{\rm{min}}$. The execution time required by the successive UE deletion increases with $R_{\rm{min}}$, and is a little higher than that of the bisection search algorithm for large $R_{\rm{min}}$, which may even decreases with $R_{\rm{min}}$.

Fig.~\ref{figcon} shows the convergence behaviour of the FOTA-based Algorithm under different rate targets, where the bisection search algorithm is employed for selecting the admitted UEs. The average numbers of admitted UEs  for different rate targets are shown in this figure. It is seen from this figure that our proposed algorithm converges rapidly and generally three iterations are sufficient for the algorithm to converge under all considered rate targets, which is appealing for practical applications. Since the number of UEs admitted for the larger $R_{\rm{min}}$ is smaller, the larger  $R_{\rm{min}}$ may not yield higher transmit power.

In Fig.~\ref{powvsrate}, we compare the performance of the FOTA-based Algorithm with that of the exhaustive search method. For the latter algorithm, if $\left| {{{\cal U}_i}} \right|\geq {{\tilde C}_{\max }}$, the algorithm checks all possible subsets of ${{{\cal U}_i}}$ with size ${{\tilde C}_{\max }}$, and chooses the one with the minimum transmit power. It is observed from Fig.~\ref{powvsrate} that our proposed algorithm achieves almost the same performance as that of the exhaustive search method, which confirms the effectiveness of our proposed algorithm. The corresponding execution time for these two algorithms is shown in Fig.~\ref{timeex}. We can observe from Fig.~\ref{timeex} that the execution time of the exhaustive search method requires much more time than the proposed FOTA-based Algorithm for the small $R_{\rm{min}}$, and almost the same for large $R_{\rm{min}}$. The reason can be explained as follows. For the case of small $R_{\rm{min}}$, more UEs can be admitted in the network, so that more RRHs will satisfy the condition $\left| {{{\cal U}_i}} \right|\geq {{\tilde C}_{\max }}$. Then the number of checking times is large, which leads to high computational complexity. However, for the case of large  $R_{\rm{min}}$, only a small number of UEs can be admitted as seen in Fig.~\ref{fig3}. Then, almost all the RRHs satisfy the fronthaul capacity constraint, and it is not necessary for the exhaustive search method to enumerate the UE-RRH associations, leading to almost the same complexity of our algorithm. Note that the time required by the proposed algorithm is within one second and the algorithm converges within five iterations as seen in Fig.~\ref{figcon}, then the execution time for each iteration of the FOTA-based Algorithm is within 0.2 second.

Now, we study the robustness  of the proposed algorithm against the following four algorithms:
\begin{enumerate}
\item Only Robust to Channel Quantization (labeled as `Robust CH Quan.'): This method only takes into account the effect of channel quantization, when designing the beamforming vectors, regardless of the channel estimation errors.
\item Only Robust to Channel Estimation Error (labeled as `Robust CH Esti.'): As the terminology suggests, this method only considers the effects of channel estimation errors, and naively treats the feedback CDI and PA as perfect. Then, the SDP method proposed in \cite{CunhuTWC} can be adopted to solve the resultant optimization problem.
\item Only Feeding back the CDI Information (labeled as `CDI FB Only'): In this method, each UE only feeds back the CDI index to the BBU pool, without considering the PA information. The $\bf{A}$ matrix derived in Appendix \ref{prooftheorem1} can be recalculated without considering the PA quantization information and the statistics of the quantization error.
\item Nonrobust Beamforming Design (labeled as `Non-robust'): Neither channel quantization errors nor channel estimation errors are considered by this algorithm  and the feedback CDI and PA are regarded as perfect.
\end{enumerate}

\begin{table}[!t]
\renewcommand{\arraystretch}{1.1}
\caption{ Power consumption and numbers of admitted UEs for various methods}\vspace{-0.4cm}
\label{tab1}
\centering
\begin{tabular}{|c|c|c|c|c|c|}
\hline
  & Proposed Robust    & Robust CH Quan.  & Robust CH Esti. & CDI FB Only & Non-robust \\
\hline
Power consumption (mW) &  167  & 130 &  158 &  149 & 114  \\
\hline
\end{tabular}\vspace{-0.8cm}
\end{table}

Table~\ref{tab1} reports the total power consumption required by the various methods for one random channel generation, where all eight UEs are admitted.  It can be seen that our proposed algorithm has the highest power consumption, since it requires more power to compensate for both the channel estimation errors and channel quantization errors. Note that the non-robust algorithm requires the least since these errors are not considered. However, it is important to observe each UE's actual achievable data rate achieved by these algorithms. Fig.~\ref{fig4} shows each UE's actual achievable data rate by all the methods. It is seen that all UEs' data requirements are satisfied by our proposed robust algorithm,  which confirms the effectiveness of our proposed algorithm. For the `Robust CH Quan.' method, all UEs' rate requirements are not fulfilled since the channel estimation errors are not considered. Hence, the channel estimation error cannot be ignored when designing the beamforming vector due to the non-negligible pilot contamination.  For `Robust CH Esti.' method, the statistics information of channel quantization  error is not considered and some UEs' actual achievable data rates are lower than the rate target, such as those of UE 2 and UE 8. It is also observed that some UEs have much higher rates, indicating that the power and spatial resources are not properly allocated by the `Robust CH Esti.' method. For the `CDI FB Only' method, the actual achievable data rates of all UEs are lower than the rate targets, and UE 2's data rate is even lower than $1\ \rm{bit/s/Hz}$. This confirms the importance of feeding back the PA information for coherent transmission. Finally, some UEs' actual achievable data rates are below the rate target by the `Non-robust' method as it naively treats the feedback CSI as the perfect. However, it is observed in Fig.~\ref{fig4} that, even with non-perfect PA feedback information, the performance of the `Non-robust' method is even better than that of the `CDI FB Only' method in terms of the number of UEs that satisfy the rate target. In summary, only our proposed algorithm is capable of maintaining the guaranteed rates for each UE, since it jointly considers the effects of channel estimation errors and channel quantization errors, which are (partially) ignored by the other algorithms.

\vspace{-0.4cm}\subsection{Larger UD-CRANs}

The following simulation results are based on the larger UD-CRAN. We investigate the effects of different system parameters on the performance of the proposed algorithm.

Since UD-CRANs will be deployed in hot spots,  where the number of UEs is high and the communication resources are limited, maximizing the number of admitted users for each time frame should be a high priority. Additionally, according to the results of Fig.~\ref{figcon}, the power consumption may not provide sufficient insights, since its value mainly depends on the number of UEs selected from Stage I. Hence, in the following, we only consider the performance in terms of the number of UEs that can be supported. For comparison, the performance of the algorithm having perfect intra-cluster CSI \cite{pan2017joint} is also simulated as a performance benchmark. The bisection based search method and the successive UE deletion method of the robust algorithm are denoted as `Robust-Bis' and `Robust-Suc', respectively, while `Perfect-intraCSI-Bis' and `Perfect-intraCSI-Suc'  represent the two methods for the case of perfect intra-cluster CSI.

\subsubsection{Impact of the number of CDI quantization bits} 
\begin{figure}
\begin{minipage}[t]{0.475\linewidth}
\centering
\includegraphics[width=2.6in]{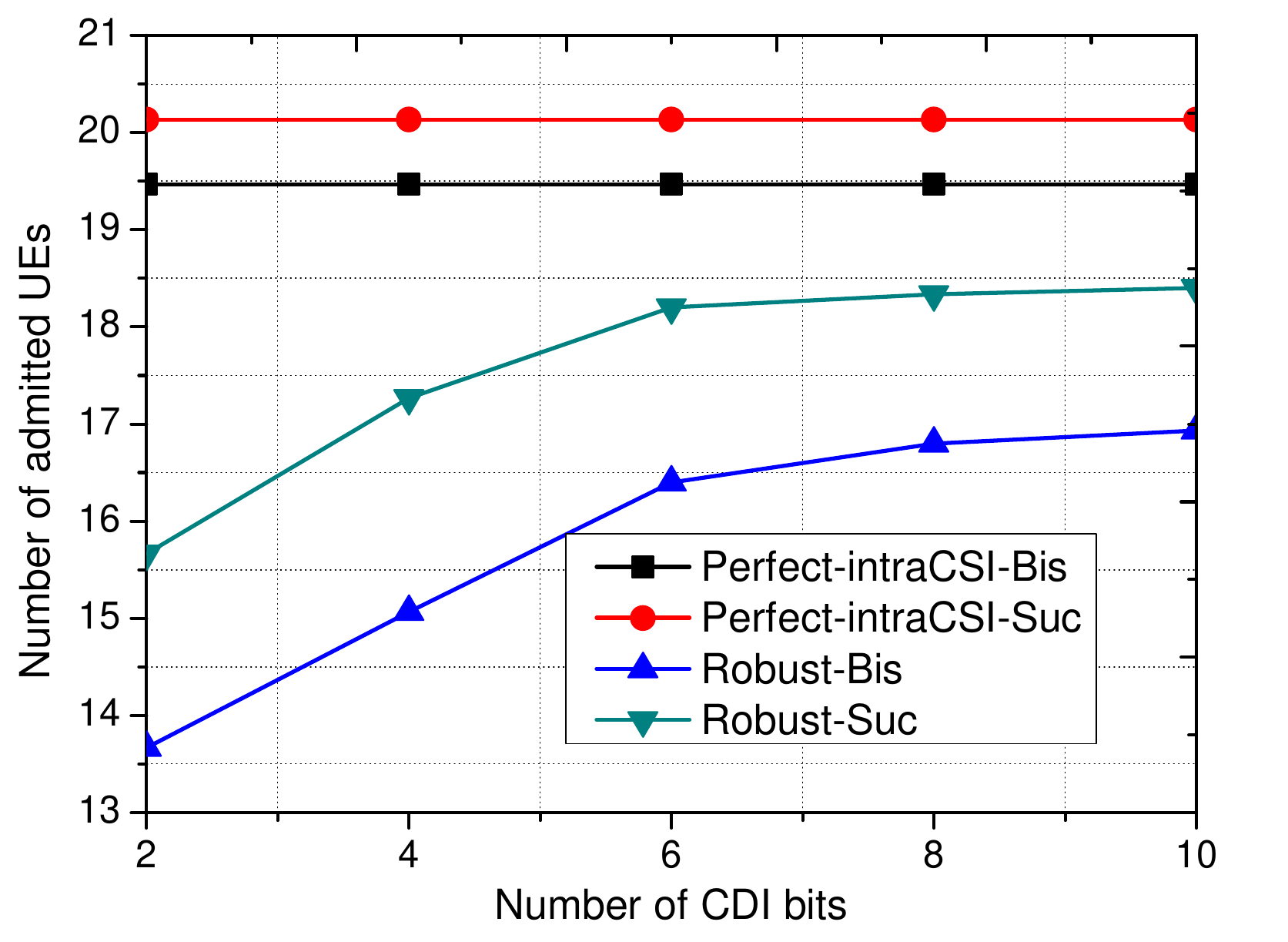}\vspace{-0.4cm}
\caption{Number of UEs admitted  by various algorithms versus CDI quantization bits $B^{\rm{CDI}}$ for a large UD-CRAN.}\vspace{-0.8cm}
\label{fig6}
\end{minipage}%
\hfill
\begin{minipage}[t]{0.475\linewidth}
\centering
\includegraphics[width=2.6in]{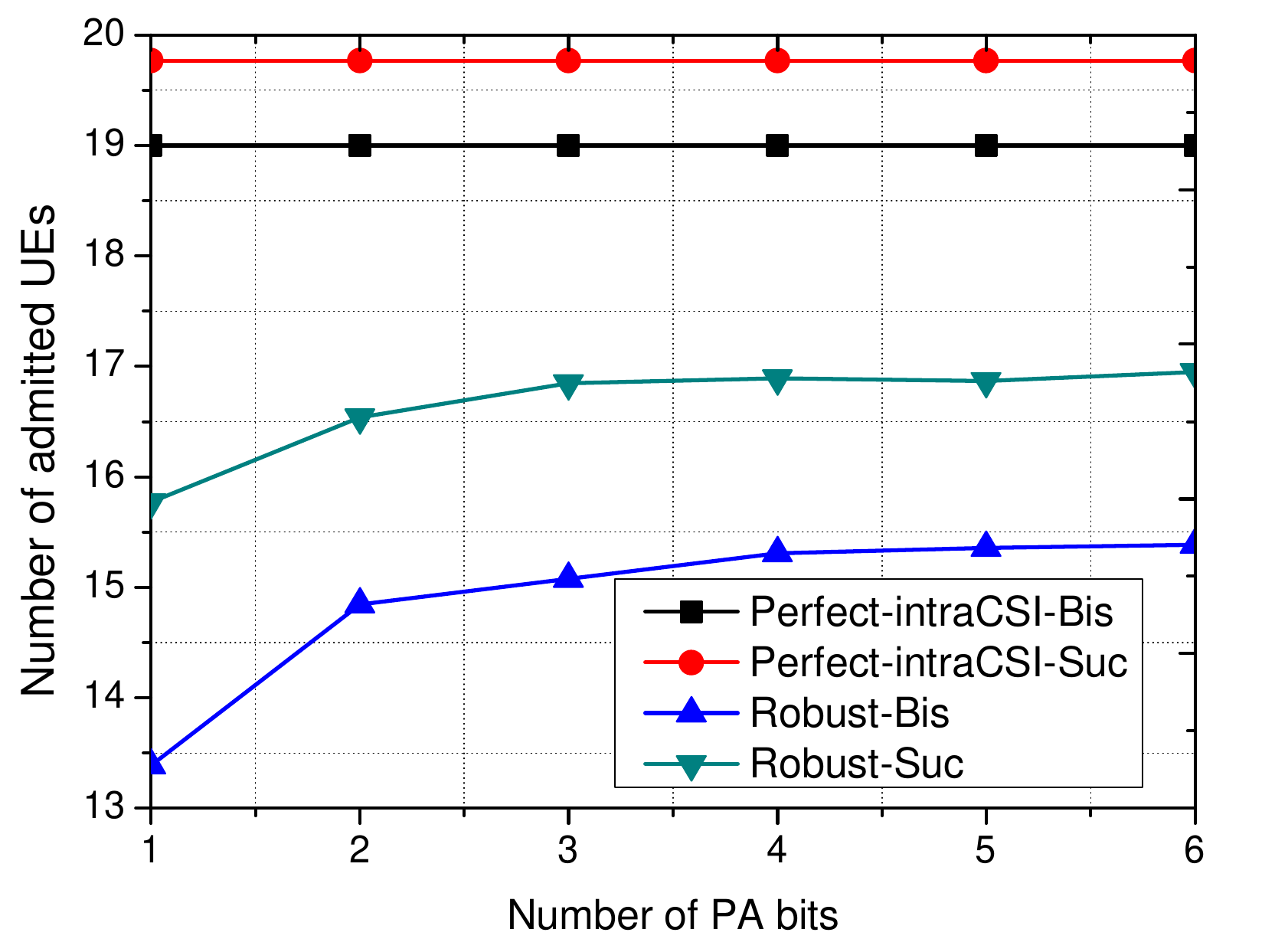}\vspace{-0.4cm}
\caption{Number of UEs admitted  by various algorithms versus PA quantization bits $B^{\rm{PA}}$ for a large UD-CRAN.}\vspace{-0.8cm}
\label{fig7}
\end{minipage}%
\hfill
\end{figure}

Fig.~\ref{fig6} illustrates the impact of CDI quantization bits $B^{{\rm{CDI}}}$ on the system performance. Similar observations can be found in Fig.~\ref{fig6} as those in Fig.~\ref{fig5}. Note that when $B^{{\rm{CDI}}}$ increases from 2 to 6, three more UEs can be admitted by the proposed robust algorithms and will not increase for $B^{{\rm{CDI}}}\geq 6$. This is due to the fact that each RRH is equipped with two antennas and a small number of CDI quantization bits are sufficient to achieve good performance. A fixed performance gap is observed between the robust algorithm and those for perfect intra-cluster CSI when $B^{{\rm{CDI}}}\geq 6$ due to the additional channel estimation error.

\subsubsection{Impact of the number of PA quantization bits} Now, we study the impact of the important system parameter $B^{\rm{PA}}$ in Fig.~\ref{fig7}. It is seen from this figure that there is a slight increase of the number of UEs admitted by the robust algorithms when $B^{\rm{PA}}$ increases from 1 to 3 and becomes saturated, when $B^{\rm{PA}}\geq 3$. This is a very inspiring result, implying that only a small number of bits is necessary for the PA quantization, which mitigates the feedback overhead, while guaranteeing good performance. In particular, even one bit used for PA quantization can achieve $90\%$ of the performance attained with perfect PA information.

\section{Conclusions}\label{conclu}
This paper provided a complete framework for dealing with the unavailability of full CSI in user-centric UD-CRANs, where only partial inter-cluster CSI and a quantized version of the intra-cluster CSI are available at the BBU pool. We derived the achievable data rate expression by exploiting the statistical characteristics of various channel uncertainties. Based on this, we developed a low-complexity robust beamforming algorithm for minimizing the total transmit power, while guaranteeing each user's rate requirement and fronthaul capacity constraints. In addition, to ensure the feasibility of the problem, a pair of low-complexity user selection algorithms are provided as well. Simulation results show that our proposed robust algorithm significantly outperforms the existing state-of-art algorithms in terms of providing the required guaranteed quality-of-service (QoS) for the users. Furthermore, extensive simulations results are provided to study the impact of different system parameters on the performance. One new important observation was made:  One bit for quantizing the each RRH's PA is enough to obtain a large proportion of the performance obtained with perfect PA information.

\numberwithin{equation}{section}
\begin{appendices}

\section{Derivations of ${\bf{A}}_{k,k}$ and ${\bf{A}}_{l,k}$}\label{prooftheorem1}
\subsection{Derivation of ${\bf{A}}_{k,k}$}
We first derive the expression of ${\bf{A}}_{k,k}$, which is equal to ${{\bf{A}}_{k,k}} = {\mathbb{E}}\left\{ {{{\bf{\hat g}}_{k,k}}{\bf{\hat g}}_{k,k}^{\rm{H}}} \right\}$.  Denote indices of ${{{\cal I}_k}}$ as ${{\cal I}_k} = \{ s_1^k, \cdots ,s_{|{{\cal I}_l}|}^k\}$. To calculate ${{\bf{A}}_{k,k}}$, we first have to calculate ${\mathbb{E}}\left\{ {{{\bf{\hat h}}_{s_i^k,k}}{\bf{\hat h}}_{s_i^k,k}^{\rm{H}}} \right\}$ and ${\mathbb{E}}\left\{ {{{\bf{\hat h}}_{s_i^k,k}}{\bf{\hat h}}_{s_l^k,k}^{\rm{H}}} \right\}$, where $i\neq l$.

The derivation of ${\mathbb{E}}\left\{ {{{\bf{\hat h}}_{s_i^k,k}}{\bf{\hat h}}_{s_i^k,k}^{\rm{H}}} \right\}$ is based on the following lemma.

\itshape \textbf{Lemma 1:}  \upshape The random vectors and variables have the following properties:
\begin{eqnarray}
{\mathbb{E}}\left\{ {{{\bf{q}}_{s_i^k,k}}{\bf{u}}_{s_i^k,k}^{\rm{H}}} \right\} &=& {\bf{0}}, \label{dewA}\\
{\mathbb{E}}\left\{ {{a_{s_i^k,k}}} \right\} &=& {2^{{B_{s_i^k,k}^{\rm{CDI}}}}}\beta \left( {{2^{{B_{s_i^k,k}^{\rm{CDI}}}}},\frac{M}{{M - 1}}} \right)\triangleq \rho_{s_i^k,k},\label{dret}\\
{\mathbb{E}}\left\{ {{{\bf{u}}_{i,k}}{\bf{u}}_{i,k}^{\rm{H}}} \right\} &=& \frac{1}{{M - 1}}\left( {{{\bf{I}}_M} - {{\bf{q}}_{s_i^k,k}}{\bf{q}}_{s_i^k,k}^{\rm{H}}} \right)\triangleq {{\bf{O}}_{s_i^k,k}},\label{hwefo}
\end{eqnarray}
where $\beta(a,b)$ is the beta function.

\itshape \textbf{Proof:}  \upshape (\ref{dewA}) follows since ${{{\bf{q}}_{s_i^k,k}}}$ is orthogonal to ${{{\bf{u}}_{s_i^k,k}}}$,  while (\ref{dret}) and (\ref{hwefo}) are from \cite{yeung2007} and \cite{zhang2009}, respectively.
\hfill $\Box$

Based on Lemma 1, we embark on deriving ${\mathbb{E}}\left\{ {{{\bf{\hat h}}_{s_i^k,k}}{\bf{\hat h}}_{s_i^k,k}^{\rm{H}}} \right\}$. According to Section-\ref{chanesti} and Section-\ref{limitfeed}, the quantized channel ${{\bf{\hat h}}_{s_i^k,k}}$ can be rewritten as
\begin{equation}\label{dewaf}
  {{\bf{\hat h}}_{s_i^k,k}} = \left\| {{{{\bf{\hat h}}}_{s_i^k,k}}} \right\|\left( {\sqrt {1 - a_{s_i^k,k}} {e^{j{\phi _{s_i^k,k}}}}{{\bf{q}}_{s_i^k,k}} + \sqrt {a_{s_i^k,k}} {{\bf{u}}_{s_i^k,k}}} \right).
\end{equation}
Then ${\mathbb{E}}\left\{ {{{\bf{\hat h}}_{s_i^k,k}}{\bf{\hat h}}_{s_i^k,k}^{\rm{H}}} \right\}$ can be derived as
\begin{eqnarray}
&&{\mathbb{E}}\left\{ {{{\bf{\hat h}}_{s_i^k,k}}{\bf{\hat h}}_{s_i^k,k}^{\rm{H}}} \right\} \nonumber\\
&=& {\mathbb{E}}\left\{ {{{\left\| {{{{\bf{\hat h}}}_{s_i^k,k}}} \right\|}^2}} \right\}\left( {{\mathbb{E}}\left\{ {\left( {1 - {a_{s_i^k,k}}} \right)} \right\}{{\bf{q}}_{s_i^k,k}}{\bf{q}}_{s_i^k,k}^{\rm{H}} + {\mathbb{E}}\left\{ {{a_{s_i^k,k}}} \right\}{\mathbb{E}}\left\{ {{{\bf{u}}_{s_i^k,k}}{\bf{u}}_{s_i^k,k}^{\rm{H}}} \right\}} \right) \label{OLHf}\\
&=& {\omega _{s_i^k,k}}M\left( {\left( {1 - {\rho _{s_i^k,k}}} \right){{\bf{q}}_{s_i^k,k}}{\bf{q}}_{s_i^k,k}^{\rm{H}} + {\rho _{s_i^k,k}}{{\bf{O}}_{s_i^k,k}}} \right),\label{joi}
\end{eqnarray}
where (\ref{dewA}) is used in (\ref{OLHf}), (\ref{dret}) and (\ref{hwefo}) are used in (\ref{joi}). Note that the PA quantization does not affect the value of  ${\mathbb{E}}\left\{ {{{\bf{\hat h}}_{s_i^k,k}}{\bf{\hat h}}_{s_i^k,k}^{\rm{H}}} \right\}$.

To calculate ${\mathbb{E}}\left\{ {{{\bf{\hat h}}_{s_i^k,k}}{\bf{\hat h}}_{s_l^k,k}^{\rm{H}}} \right\}$ with $i\neq l$, the following lemma should be employed.

\itshape \textbf{Lemma 2:}  \upshape The random vectors and variables in (\ref{dewaf}) have the following properties:
\begin{eqnarray}
{\mathbb{E}}\left\{ {{e^{j{\phi _{s_i^k,k}}}}} \right\} &=&{e^{j{{\hat \phi }_{s_i^k,k}}}}\frac{{{2^{B_{s_i^k,k}^{{\rm{PA}}}}}}}{\pi }\sin \left( {\frac{\pi }{{{2^{B_{s_i^k,k}^{{\rm{PA}}}}}}}} \right)\triangleq {e^{j{{\hat \phi }_{s_i^k,k}}}}\xi_{s_i^k,k}  \label{hurelfu}\\
{\mathbb{E}}\left\{ {\left\| {{{{\bf{\hat h}}}_{s_i^k,k}}} \right\|} \right\} &=& \frac{{\sqrt {{\omega _{s_i^k,k}}} \Gamma \left( {M + \frac{1}{2}} \right)}}{{\Gamma \left( M \right)}}\triangleq \varsigma_{s_i^k,k} \label{halif} \\
{\mathbb{E}}\left\{ {\sqrt {1 - {a_{s_i^k,k}}} } \right\} &=& \sum\limits_{m = 1}^{{2^{B_{s_i^k,k}^{{\rm{CDI}}}}}} {C_{{2^{B_{s_i^k,k}^{{\rm{CDI}}}}}}^m} {\left( { - 1} \right)^{m + 1}}m(M - 1)\beta \left( {m\left( {M - 1} \right),\frac{3}{2}} \right)\triangleq {\Omega _{s_i^k,k}},\label{relur}
\end{eqnarray}
where  $C_n^r = \frac{{n!}}{{r!(n - r)!}}$.

\itshape \textbf{Proof:}  \upshape We first prove equality (\ref{hurelfu}). Specifically, we have
\begin{eqnarray}
{\mathbb{E}}\left\{ {{e^{j{\phi _{s_i^k,k}}}}} \right\} &=& {e^{j{{\hat \phi }_{s_i^k,k}}}}{\mathbb{E}}\left\{ {{e^{j{{\tilde \phi }_{s_i^k,k}}}}} \right\}= {e^{j{{\hat \phi }_{s_i^k,k}}}}\int_{ - \frac{\pi }{{{2^{B_{s_i^k,k}^{{\rm{PA}}}}}}}}^{\frac{\pi }{{{2^{B_{s_i^k,k}^{{\rm{PA}}}}}}}} {{e^{jx}}\frac{{{2^{B_{s_i^k,k}^{{\rm{PA}}}}}}}{{2\pi }}dx}\nonumber\\
  &=& {e^{j{{\hat \phi }_{s_i^k,k}}}}\int_{ - \frac{\pi }{{{2^{B_{s_i^k,k}^{{\rm{PA}}}}}}}}^{\frac{\pi }{{{2^{B_{s_i^k,k}^{{\rm{PA}}}}}}}} {\left( {\cos x + j\sin x} \right)\frac{{{2^{B_{s_i^k,k}^{{\rm{PA}}}}}}}{{2\pi }}dx}\nonumber\\
  &=&{e^{j{{\hat \phi }_{s_i^k,k}}}}\frac{{{2^{B_{s_i^k,k}^{{\rm{PA}}}}}}}{\pi }\sin \left( {\frac{\pi }{{{2^{B_{s_i^k,k}^{{\rm{PA}}}}}}}} \right).\nonumber
\end{eqnarray}

For (\ref{halif}), let us define ${{{\bf{\mathord{\buildrel{\lower3pt\hbox{$\scriptscriptstyle\frown$}}
\over h} }}}_{s_i^k,k}} \buildrel \Delta \over = {{{{{\bf{\hat h}}}_{s_i^k,k}}} \mathord{\left/
 {\vphantom {{{{{\bf{\hat h}}}_{s_i^k,k}}} {\sqrt {{\omega _{s_i^k,k}}} }}} \right.
 \kern-\nulldelimiterspace} {\sqrt {{\omega _{s_i^k,k}}} }}$. Then, $\left\| {{{{\bf{\mathord{\buildrel{\lower3pt\hbox{$\scriptscriptstyle\frown$}}
\over h} }}}_{s_i^k,k}}} \right\|$ obeys a scaled  (by a factor of ${1 \mathord{\left/
 {\vphantom {1 {\sqrt 2 }}} \right.
 \kern-\nulldelimiterspace} {\sqrt 2 }}$) chi distribution with $2M$ degrees of freedom. Hence, ${\mathbb{E}}\left\{ {\left\| {{{{\bf{\mathord{\buildrel{\lower3pt\hbox{$\scriptscriptstyle\frown$}}
\over h} }}}_{s_i^k,k}}} \right\|} \right\} = \frac{{\Gamma (M + \frac{1}{2})}}{{\Gamma (M)}}$ \cite{Jose2011} and (\ref{halif}) is proved.

Finally, we embark on proving (\ref{relur}). Define $\nu \triangleq 1-a_{s_i^k,k}$. According to Lemma 1 in \cite{yeung2007}, the probability density function of $\nu$ with $M$ antennas using a ${{2^{B}}}$ RVQ codebook is given by
\begin{equation}\label{dewa}
  {f_\upsilon }(\upsilon ) = \sum\limits_{i = 1}^{{2^B}} {C_{{2^B}}^i{{\left( { - 1} \right)}^{i + 1}}i(M - 1){{\left( {1 - \upsilon } \right)}^{i\left( {M - 1} \right) - 1}}}.
\end{equation}
Define $x \triangleq \sqrt \upsilon   \in \left[ {0,1} \right]$, then the Jacobian of the transformation is given by $J = \frac{{dx}}{{d\upsilon }} = \frac{1}{2}{\upsilon ^{ - \frac{1}{2}}}$. Hence, the PDF of $x$ is given by
\begin{equation}\label{dadeeda}
  {f_x}(x) = 2\sum\limits_{i = 1}^{{2^B}} {C_{{2^B}}^i{{\left( { - 1} \right)}^{i + 1}}i(M - 1){{\left( {1 - {x^2}} \right)}^{i\left( {M - 1} \right) - 1}}} x.
\end{equation}
Then the expectation of $x$ is given by
\begin{eqnarray}
{\mathbb{E}}\{ x\}  &=& \int_0^1 x {f_x}(x)dx\nonumber\\
&=& 2\sum\limits_{m = 1}^{{2^B}} {C_{{2^B}}^m{{\left( { - 1} \right)}^{m + 1}}m(M - 1)\int_0^1 {{{\left( {1 - {x^2}} \right)}^{m\left( {M - 1} \right) - 1}}{x^2}dx} } \nonumber\\
&=& \sum\limits_{m = 1}^{{2^B}} {C_{{2^B}}^m{{\left( { - 1} \right)}^{m + 1}}m(M - 1)\beta \left( {m(M - 1),\frac{3}{2}} \right)}\nonumber
\end{eqnarray}
where the last equality is due to the fact that the beta function can be represented as \cite{gupta2004handbook,Jinadl}
\begin{equation}\label{zef}
 \beta \left( {c,\frac{a}{b}} \right) = b\int_0^1 {{x^{a - 1}}{{\left( {1 - {x^b}} \right)}^{c - 1}}dx}
\end{equation}
with $a=3$, $b=2$ and $c=m(M-1)$. Hence, (\ref{relur}) follows with $B=B_{s_i^k,k}^{{\rm{CDI}}}$. \hfill $\Box$

Based on Lemma 2, ${\mathbb{E}}\left\{ {{{\bf{\hat h}}_{s_i^k,k}}{\bf{\hat h}}_{s_l^k,k}^{\rm{H}}} \right\}, i \ne l$, can be calculated as
\begin{eqnarray}
{\mathbb{E}}\left\{ {{{\bf{\hat h}}_{s_i^k,k}}{\bf{\hat h}}_{s_l^k,k}^{\rm{H}}} \right\} &=& {\mathbb{E}}\left\{ {{{\bf{\hat h}}_{s_i^k,k}}} \right\}{\mathbb{E}}\left\{ {{\bf{\hat h}}_{s_l^k,k}^{\rm{H}}} \right\}\label{furelh}\\
&=& {\varsigma _{s_i^k,k}}{\varsigma _{s_l^k,k}}{\Omega _{s_i^k,k}}{\Omega _{s_l^k,k}}{\xi _{s_i^k,k}}{\xi _{s_l^k,k}}{e^{j\left( {{{\hat \phi }_{s_i^k,k}} - {{\hat \phi }_{s_l^k,k}}} \right)}} {{\bf{q}}_{s_i^k,k}}{\bf{q}}_{s_l^k,k}^{\rm{H}}\label{pooko}
\end{eqnarray}
where (\ref{furelh}) follows since ${\bf{h}}_{s_i^k,k}$ is independent of ${\bf{h}}_{s_l^k,k}$, (\ref{pooko}) follows since ${\mathbb{E}}\left\{ {{{\bf{u}}_{s_i^k,k}}} \right\} = {{\bf{0}}_M}$, and Lemma 2 is used. \hfill $\Box$

Based on the above results, ${\bf{A}}_{k,k}$ is given by
\begin{equation}\label{akk}
{\bf{A}}_{k,k} = \left[ {\begin{array}{*{20}{c}}
{{{\left( {{\bf{A}}_{k,k}} \right)}_{1,1}}}& \cdots &{{{\left( {{\bf{A}}_{k,k}} \right)}_{1,|{{\cal I}_k}|}}}\\
 \vdots & \ddots & \vdots \\
{{{\left( {{\bf{A}}_{k,k}} \right)}_{|{{\cal I}_k}|,1}}}& \cdots &{{{\left( {{\bf{A}}_{k,k}} \right)}_{|{{\cal I}_k}|,|{{\cal I}_k}|}}}
\end{array}} \right],
\end{equation}
where  ${\left( {{\bf{A}}_{k,k}} \right)_{i,l}} \in {{\mathbb{C}}^{M \times M}},i,l \in 1, \cdots ,|{{\cal I}_k}|$  is the block matrix of ${\bf{A}}_{k,k}$ at the $i$th row and $l$th column. If $i=l$, then ${\left( {{{\bf{A}}_{k,k}}} \right)_{i,l}} = {\mathbb{E}}\left\{ {{{\bf{\hat h}}_{s_i^k,k}}{\bf{\hat h}}_{s_i^k,k}^{\rm{H}}} \right\}$. Otherwise, ${\left( {{{\bf{A}}_{k,k}}} \right)_{i,l}} = {\mathbb{E}}\left\{ {{{\bf{\hat h}}_{s_i^k,k}}{\bf{\hat h}}_{s_l^k,k}^{\rm{H}}} \right\}$.

\vspace{-0.5cm}\subsection{Derivation of ${\bf{A}}_{l,k}$}

Let us denote the indices of ${{{\cal I}_l}}$ by ${{\cal I}_l} = \{ s_1^l, \cdots ,s_{|{{\cal I}_l}|}^l\}$. Then ${\bf{A}}_{l,k}$ can be represented as
\begin{equation}\label{jiofuh}
{\bf{A}}_{l,k} = \left[ {\begin{array}{*{20}{c}}
{{{\left( {{\bf{A}}_{l,k}} \right)}_{1,1}}}& \cdots &{{{\left( {{\bf{A}}_{l,k}} \right)}_{1,|{{\cal I}_l}|}}}\\
 \vdots & \ddots & \vdots \\
{{{\left( {{\bf{A}}_{l,k}} \right)}_{|{{\cal I}_l}|,1}}}& \cdots &{{{\left( {{\bf{A}}_{l,k}} \right)}_{|{{\cal I}_l}|,|{{\cal I}_l}|}}}
\end{array}} \right],
\end{equation}
where ${\left( {{{\bf{A}}_{l,k}}} \right)_{i,m}} = {\mathbb{E}}\left\{ {{{\bf{h}}_{s_i^l,k}}{\bf{h}}_{s_m^l,k}^{\rm{H}}} \right\}$.
To derive ${\mathbb{E}}\left\{ {{{\bf{h}}_{s_i^l,k}}{\bf{h}}_{s_m^l,k}^{\rm{H}}} \right\}$, we should  discuss four cases: 1) $s_i^l,s_m^l \in {{\cal I}_k},i \ne m$; 2) $s_i^l,s_m^l \in {{\cal I}_k},i = m$; 3) $s_i^l,s_m^l \notin {{\cal I}_k},i = m$; 4) $s_i^l \notin {\cal I}_k$ or $s_m^l \notin {\cal I}_k$, $i\ne m$.

For Case 1), both RRH $s_i^l$ and RRH $s_m^l$ belong to UE $k$'s cluster ${\cal I}_k$, but they are not the same RRH.
Then, ${{\bf{h}}_{s_i^l,k}}$ can be rewritten as
\begin{equation}\label{dwefdseds}
 {{\bf{h}}_{s_i^l,k}} = \left\| {{{{\bf{\hat h}}}_{s_i^l,k}}} \right\|\left( {\sqrt {1 - a_{s_i^l,k}} {e^{j{\phi _{s_i^l,k}}}}{{\bf{q}}_{s_i^l,k}} + \sqrt {a_{s_i^l,k}} {{\bf{u}}_{s_i^l,k}}} \right) + {{\bf{e}}_{s_i^l,k}}.
\end{equation}
By exploiting the fact that ${\mathbb{E}}\left\{ {{{\bf{u}}_{s_i^l,k}}} \right\} = {{\bf{0}}_M}$, ${\mathbb{E}}\left\{ {{{\bf{e}}_{s_i^l,k}}} \right\} = {{\bf{0}}_M}$ and Lemma 2, ${\mathbb{E}}\left\{ {{{\bf{h}}_{s_i^l,k}}{\bf{h}}_{s_m^l,k}^{\rm{H}}} \right\}$ can be derived similarly to ${\mathbb{E}}\left\{ {{{\bf{\hat h}}_{i,k}}{\bf{\hat h}}_{l,k}^{\rm{H}}} \right\}$ in (\ref{pooko}), which is given by
\begin{equation}\label{frese}
{\mathbb{E}}\left\{ {{{\bf{h}}_{s_i^l,k}}{\bf{h}}_{s_m^l,k}^{\rm{H}}} \right\}={\varsigma _{s_i^l,k}}{\varsigma _{s_m^l,k}}{\Omega _{s_i^l,k}}{\Omega _{s_m^l,k}}{\xi _{s_i^l,k}}{\xi _{s_m^l,k}}{e^{j\left( {{{\hat \phi }_{s_i^l,k}} - {{\hat \phi }_{s_m^l,k}}} \right)}} {{\bf{q}}_{s_i^l,k}}{\bf{q}}_{s_m^l,k}^{\rm{H}}.
\end{equation}

 For Case 2),  RRH $s_i^l$ and RRH $s_m^l$ represent the same RRH belonging to ${\cal I}_k$. By using (\ref{dwefdseds}) and the facts that ${\mathbb{E}}\left\{ {{{\bf{q}}_{s_i^l,k}}{\bf{e}}_{s_i^l,k}^{\rm{H}}} \right\} ={\bf{0}}$, ${\mathbb{E}}\left\{ {{{\bf{u}}_{s_i^l,k}}{\bf{e}}_{s_i^l,k}^{\rm{H}}} \right\} = {\bf{0}}$ and Lemma 1, ${\mathbb{E}}\left\{ {{{\bf{h}}_{s_i^l,k}}{\bf{h}}_{s_m^l,k}^{\rm{H}}} \right\}$ can be derived  similarly to  ${\mathbb{E}}\left\{ {{{\bf{\hat h}}_{i,k}}{\bf{\hat h}}_{i,k}^{\rm{H}}} \right\}$ in (\ref{joi}), which is given by
 \begin{equation}\label{dewad}
 {\mathbb{E}}\left\{ {{{\bf{h}}_{s_i^l,k}}{\bf{h}}_{s_m^l,k}^{\rm{H}}} \right\}={\omega _{s_i^l,k}}M\left[ {\left( {1 - {\rho _{s_i^l,k}}} \right){{\bf{q}}_{s_i^l,k}}{\bf{q}}_{s_i^l,k}^{\rm{H}} + {\rho _{s_i^l,k}}{{\bf{O}}_{s_i^l,k}}} \right] + {\delta_{s_i^l,k}}{{\bf{I}}_M},
 \end{equation}
 where $\delta _{s_i^l,k}$ is the channel estimation error given in (\ref{degrag}).

 For Case 3), both RRHs represent the same RRH that is not in UE $k$'s cluster ${\cal I}_k$. Then, ${\mathbb{E}}\left\{ {{{\bf{h}}_{s_i^l,k}}{\bf{h}}_{s_m^l,k}^{\rm{H}}} \right\}$ is given by ${\mathbb{E}}\left\{ {{{\bf{h}}_{s_i^l,k}}{\bf{h}}_{s_m^l,k}^{\rm{H}}} \right\} = {\alpha _{s_i^l,k}}{{\bf{I}}_M}$, since we have assumed that only large-scale fading gains are available for the out-cluster RRHs  in the BBU pool.

 For the latter case, it can be readily shown that ${\mathbb{E}}\left\{ {{{\bf{h}}_{s_i^l,k}}{\bf{h}}_{s_m^l,k}^{\rm{H}}} \right\} = {{\bf{0}}_M}$, since at least one RRH does not belong to ${\cal I}_k$ and they are not the same RRH.

\section{Derivation of Inequality (\ref{ehdliu})}\label{hfiureah}
Before deriving (\ref{ehdliu}), we first introduce the concept of $\cal T$-transform. For a complex vector ${\bf{x}}\in{\cal \mathbb C}^{n}$ and a complex matrix ${\bf{X}}\in{\mathbb C}^{n \times m}$, a one-to-one mapping function of $ {\mathbb C}^{n} \to {\mathbb R}^{2n}$ and $ {\mathbb C}^{n\times m} \to {\mathbb R}^{2n \times 2m}$, is defined as
\begin{equation}\label{jfierjuir}
 \cal T({\bf{x}}) = \left[ \begin{array}{l}
{\mathop{\rm Re}\nolimits} ({\bf{x}})\\
{\mathop{\rm Im}\nolimits} ({\bf{x}})
\end{array} \right],T({\bf{X}}) = \left[ {\begin{array}{*{20}{c}}
{{\mathop{\rm Re}\nolimits} ({\bf{x}})}&{ - {\mathop{\rm Im}\nolimits} ({\bf{x}})}\\
{{\mathop{\rm Im}\nolimits} ({\bf{x}})}&{{\mathop{\rm Re}\nolimits} ({\bf{x}})}
\end{array}} \right].
\end{equation}
The $\cal T$-transform establishes the relationship between complex-valued vectors or matrices and their  counterparts, which facilitates the derivation of the first-order Taylor expansion  for a complex-valued function.

Let us define ${\bf{\hat x}} \buildrel \Delta \over =\cal T({\bf{x}})$ and ${\bf{\hat X}} \buildrel \Delta \over = \cal T({\bf{X}})$ as the $\cal T$-transform results of the complex-valued vector ${\bf{x}}$ and matrix ${\bf{X}}$, respectively. Then we have the following two properties for the $\cal T$-transform \cite{telatar1999capacity}:
\begin{eqnarray}
{\bf{y = Ax}} &\Leftrightarrow& {\bf{\hat y = \hat A\hat x}},\label{deweUIIU}\\
{\mathop{\rm Re}\nolimits} \left( {{{\bf{x}}^{\rm{H}}}{\bf{y}}} \right) &=& {{{\bf{\hat x}}}^{\rm{T}}}{\bf{\hat y}}.\label{juilufhlare}
\end{eqnarray}
Based on the above results,   ${\bf{w}}_k^{\rm{H}}{{\bf{A}}_{k,k}}{{\bf{w}}_k}$ can be equivalently written for the real-valued vector ${{{\bf{\hat w}}}_k}$ and for the matrix ${{\bf{A}}_{k,k}}$ as
\begin{equation}\label{ehwyureh}
  {\bf{w}}_k^{\rm{H}}{{\bf{A}}_{k,k}}{{\bf{w}}_k} \mathop = \limits^{(a)} {\mathop{\rm Re}\nolimits} \left( {{\bf{w}}_k^{\rm{H}}{{\bf{A}}_{k,k}}{{\bf{w}}_k}} \right)\mathop = \limits^{(b)} {\bf{\hat w}}_k^{\rm{T}}{\cal T}\left( {{{\bf{A}}_{k,k}}{{\bf{w}}_k}} \right) \mathop = \limits^{(c)} {\bf{\hat w}}_k^{\rm{T}}{{{\bf{\hat A}}}_{k,k}}{{{\bf{\hat w}}}_k}\triangleq g({\bf{\hat w}}_k),
\end{equation}
where (a) follows since ${\bf{w}}_k^{\rm{H}}{{\bf{A}}_{k,k}}{{\bf{w}}_k}$ is a real value, (b) follows by using (\ref{juilufhlare}) and (c) follows by using (\ref{deweUIIU}). Since $g({\bf{\hat w}}_k)$ is a convex function of ${\bf{\hat w}}_k$,  we have
\begin{eqnarray}
{\bf{w}}_k^{\rm{H}}{{\bf{A}}_{k,k}}{{\bf{w}}_k}&=& g({{{\bf{\hat w}}}_k})\\
 &\ge& g[{{{\bf{\hat w}}}_k}(t)] + \nabla g{[{{{\bf{\hat w}}}_k}(t)]^{\rm{H}}}\left[ {{{{\bf{\hat w}}}_k} - {{{\bf{\hat w}}}_k}(t)} \right]\\
 &=& g[{{{\bf{\hat w}}}_k}(t)] + 2{\bf{\hat w}}_k^{\rm{H}}{{{\bf{\hat A}}}_{k,k}}\left[ {{{{\bf{\hat w}}}_k} - {{{\bf{\hat w}}}_k}(t)} \right]\label{dewafroiue}\\
 &=& {\bf{w}}_k^{\rm{H}}(t){{\bf{A}}_{k,k}}{{\bf{w}}_k}(t) + 2{\mathop{\rm Re}\nolimits} \left[ {{\bf{w}}_k^{\rm{H}}(t){{\bf{A}}_{k,k}}\left( {{{\bf{w}}_k} - {{\bf{w}}_k}(t)} \right)} \right],\label{hfuiariu}
\end{eqnarray}
where (\ref{dewafroiue}) follows since $\nabla g({{{\bf{\hat w}}}_k}(t)) = 2{{{\bf{\hat A}}}_{k,k}}{{{\bf{\hat w}}}_k}(t)$, (\ref{deweUIIU}) and (\ref{juilufhlare}) are used in (\ref{hfuiariu}) similarly to (\ref{ehwyureh}). Hence, the proof is complete.
\section{Proof of Slater's Condition of Problem  ${\cal P}_6$}\label{fjreeo}

Without loss of generality, we consider Problem ${\cal P}_6$ in the first iteration of Algorithm  \ref{algorithmiterFOTA}, i.e., $t=1$. As explained in Subsection \ref{fjjfrijojoy}, the beamforming obtained from solving Problem ${\cal P}_7$ in Section \ref{userselec} (denoted as $\bf{w}^\star$) is set as the initial beamforming in Algorithm \ref{algorithmiterFOTA}, i.e., ${\bf{w}}(0)=\bf{w}^\star$. Hence, ${\bf{w}}^\star$ is a feasible solution to Problem ${\cal P}_6$.

The idea of the proof is to construct a new set of beam-vectors from $\bf{w}^\star$ such that Constraints C2, C7 and C8 in Problem ${\cal P}_6$ hold with strict inequalities \cite{boyd2004convex}. As stated at the end of Section \ref{userselec}, the iterative algorithm to solve Problem ${\cal P}_7$ will terminate once the intermediate solutions of  $\{\varphi_k\}_{ k\in {\cal U}}$ are all equal to zero. Hence, the data rates achieved by some UEs with the obtained solution $\bf{w}^\star$ will be strictly  larger than its minimum rate requirements. We assume that UE $k$ is one of those UEs, which satisfies
\begin{equation}\label{arfgtg}
 2{\rm{Re}}\left( {{\bf{w}}_k^{\rm{H}}(0){{\bf{A}}_{k,k}}{\bf{w}}_k^\star} \right) - {\zeta _k}(0) >{\eta _{k,\min }}\left( {{\bf{w}}_k^{{\rm{\star H}}}{{\bf{E}}_{k,k}}{\bf{w}}_k^\star + \sum\nolimits_{l \ne k,l \in {\cal U}} {{\bf{w}}_l^{{\rm{\star H}}}{{\bf{A}}_{l,k}}{\bf{w}}_l^\star + \sigma _k^2} } \right).
\end{equation}
We then scale UE $k$'s beam-vector by a constant $0<\sqrt \chi_k<1$ and denote the new beam-vector as ${\bf{w}}_k^\#  = \sqrt \chi_k  {\bf{w}}_k^\star$. One should find such a $\chi_k$ that satisfies the following inequality:
\begin{equation}\label{atregtg}
 2{\rm{Re}}\left( {{\bf{w}}_k^{\rm{H}}(0){{\bf{A}}_{k,k}}{\bf{w}}_k^\#} \right) - {\zeta _k}(0) >{\eta _{k,\min }}\left( {{\bf{w}}_k^{{\rm{\# H}}}{{\bf{E}}_{k,k}}{\bf{w}}_k^\# + \sum\nolimits_{l \ne k,l \in {\cal U}} {{\bf{w}}_l^{{\rm{\star H}}}{{\bf{A}}_{l,k}}{\bf{w}}_l^\star + \sigma _k^2} } \right).
\end{equation}
By substituting the expressions of ${\zeta _k}(0)$ and ${\bf{w}}_k^\#$ into (\ref{atregtg}), we have
\begin{equation}\label{atrdewdretg}
 {\bf{w}}_k^{{\rm{\star H}}}{{\bf{A}}_{k,k}}{\bf{w}}_k^\star\! >\! {\eta _{k,\min }}\left[ {\frac{{{\chi _k}}}{{2\sqrt {{\chi _k}}  - 1}}{\bf{w}}_k^{{\rm{\star H}}}{{\bf{E}}_{k,k}}{\bf{w}}_k^\star \!+\! \frac{1}{{2\sqrt {{\chi _k}}  - 1}}\left( {\sum\limits_{l \ne k,l \in {\cal U}} {{\bf{w}}_l^{{\rm{\star H}}}{{\bf{A}}_{l,k}}{\bf{w}}_k^\star}  \!+\!\sigma _k^2} \right)} \right].
\end{equation}
Hence, when $\frac{1}{4} < {\chi _k} < 1$, $0<{\frac{{{\chi _k}}}{{2\sqrt {{\chi _k}}  - 1}}}<1$ and $0<{\frac{1}{{2\sqrt {{\chi _k}}  - 1}}}<1$ hold. Then, one can always find a $\chi _k$ that is very close to one such that (\ref{atrdewdretg}) is satisfied.

By keeping the beam-vectors of all other UEs fixed, we immediately have
\begin{eqnarray}
&&2{\rm{Re}}\left( {{\bf{w}}_l^{\rm{H}}(0){{\bf{A}}_{l,l}}{\bf{w}}_l^\star} \right) - {\zeta _l}(0)>\nonumber\\
&&{\eta _{l,\min }}\left( {{\bf{w}}_l^{ \star {\rm{H}}}{{\bf{E}}_{l,l}}{\bf{w}}_l^ \star  + \sum\limits_{j \ne l,k,j \in {\cal U}}  {\bf{w}}_j^{ \star {\rm{H}}}{{\bf{A}}_{j,l}}{\bf{w}}_j^ \star  + {\chi _k}{\bf{w}}_k^{ \star {\rm{H}}}{{\bf{A}}_{k,l}}{\bf{w}}_k^ \star  + \sigma _l^2} \right),\forall l \ne k,l \in {\cal U}.\label{refr}
\end{eqnarray}
Hence, Constraint C8 in Problem ${\cal P}_6$ with the new set of beam-vectors $\{{\bf{w}}_k^\#, {\bf{w}}_l^\star,\forall l\neq k, \}$ hold with strict inequality for all UEs.

The remaining task is to prove that Constraint C2 and C6 hold with strict inequality. Unfortunately, with the new beam-vectors $\{{\bf{w}}_k^\#, {\bf{w}}_l^\star,\forall l\neq k, \}$, we only guarantee the following strict inequalities corresponding to the RRHs in ${\cal I}_k$:
\begin{eqnarray}
\sum\nolimits_{l \ne k,l \in {{\cal U}_i}} {{{\left\| {{\bf{w}}_{i,l}^\star} \right\|}^2}}  + {\chi _k}{\left\| {{\bf{w}}_{i,k}^ \star } \right\|^2} < {P_{i,\max }},i \in {{\cal I}_k},\label{frgt}\\
\sum\nolimits_{l \ne k,l \in {{\cal U}_i}} {{\tau _{i,l}}(0){{\left\| {{\bf{w}}_{i,l}^ \star } \right\|}^2}}  + {\tau _{i,k}}(0){\chi _k}{\left\| {{\bf{w}}_{i,k}^ \star } \right\|^2} < {{\tilde C}_i}(0),i \in {{\cal I}_k}.\label{nkyk}
\end{eqnarray}

To deal with this issue, we randomly select one RRH from ${\cal I}\backslash {{\cal I}_k}$, say RRH $i$. Then, randomly select one UE served by RRH $i$, say UE $l$. We  perform the same scaling operation as UE $k$ for UE $l$, i.e., ${\bf{w}}_l^\#  = \sqrt \chi_l  {\bf{w}}_l^\star$. One can find a $\chi _l$ ($\frac{1}{4} < {\chi _l} < 1$) such that
\begin{eqnarray}
&&2{\rm{Re}}\left( {{\bf{w}}_l^{\rm{H}}(0){{\bf{A}}_{l,l}}{\bf{w}}_l^\#} \right) - {\zeta _l}(0)>\nonumber\\
&&{\eta _{l,\min }}\left( {{\bf{w}}_l^{ \# {\rm{H}}}{{\bf{E}}_{l,l}}{\bf{w}}_l^ \#  + \sum\limits_{j \ne l,k,j \in {\cal U}}  {\bf{w}}_j^{ \star {\rm{H}}}{{\bf{A}}_{j,l}}{\bf{w}}_j^ \star  + {\chi _k}{\bf{w}}_k^{ \star {\rm{H}}}{{\bf{A}}_{k,l}}{\bf{w}}_k^ \star  + \sigma _l^2} \right).\label{rdcsdffr}
\end{eqnarray}
Obviously, with the new set of beam-vectors $\{{\bf{w}}_k^\#, {\bf{w}}_l^\#, {\bf{w}}_j^\star,\forall j\neq k, j\neq l \}$, Constraint C8 corresponding to the other UEs hold with strict inequality. Then, Constraint C2 and C6 corresponding to the RRHs in ${\cal I}_l$ hold with strict inequality. Repeat this step until Constraint C2 and C6 of all the RRHs in $\cal I$ hold with strict inequality. Then, the final constructed set of beam-vectors remain in the interior of the feasible region of Problem ${\cal P}_6$. Hence, according to Page 226 in \cite{boyd2004convex}, the Slater's condition of Problem ${\cal P}_6$  is satisfied. For Problem ${\cal P}_6$ in the subsequent iterations of Algorithm  \ref{algorithmiterFOTA}, the similar proof applies.

\section{The equivalence between Problem  ${\cal P}_1$ and Problem ${\cal P}_7$}\label{hfdewafreh}

Denote the optimal solution of Problem  ${\cal P}_1$ and Problem ${\cal P}_7$ as $\left\{ {{\cal U}^\star,{\bf{w}}^\star} \right\}$ and $\left\{ {{\cal U}^\#,{\bf{w}}^\#} \right\}$, respectively. We first prove that the optimal solution of Problem  ${\cal P}_1$ is feasible for Problem ${\cal P}_7$. It is obvious that $\left\{ {{\cal U}^\star,{\bf{w}}^\star} \right\}$ is feasible for Constraints C2 and C4 of Problem ${\cal P}_7$ since Constraint C4 is the equivalent transformation of Constraint C1 in Problem  ${\cal P}_1$. Now we show that $\left\{ {{\cal U}^\star,{\bf{w}}^\star} \right\}$ is also feasible for Problem ${\cal P}_7$. Specifically, we have the following chain inequalities:
\begin{equation}\label{dewafref}
 \sum\nolimits_{k \in {\cal U}_i^\star} {\varepsilon \left( {{{\left\| {{\bf{w}}_{i,k}^\star} \right\|}^2}} \right){R_{k,\min }}}  \le \sum\nolimits_{k \in {\cal U}_i^\star} {\varepsilon \left( {{{\left\| {{\bf{w}}_{i,k}^\star} \right\|}^2}} \right)r_k^\star}  \le {C_{i,\max }},\forall i \in {\cal I},
\end{equation}
where $r_k^\star$ is obtained by substituting ${\bf{w}}^\star$ into (\ref{datarateUEk}). Then, $\left\{ {{\cal U}^\star,{\bf{w}}^\star} \right\}$ satisfies Constraint C5 of Problem ${\cal P}_7$. Hence, $\left\{ {{\cal U}^\star,{\bf{w}}^\star} \right\}$  is  feasible for Problem ${\cal P}_7$.

For Problem ${\cal P}_7$,  if $r_k^\#  = {R_{k,\min }},\forall k$, where $r_k^\#$ is obtained by substituting ${\bf{w}}^\#$ into (\ref{datarateUEk}), then we have
 \begin{equation}\label{freooijt}
   \sum\nolimits_{k \in {\cal U}_i^\# } {\varepsilon \left( {{{\left\| {{\bf{w}}_{i,k}^\# } \right\|}^2}} \right)r_k^\# }  = \sum\nolimits_{k \in {\cal U}_i^\# } {\varepsilon \left( {{{\left\| {{\bf{w}}_{i,k}^\# } \right\|}^2}} \right){R_{k,\min }}}  \le {C_{i,\max }},\forall i \in {\cal I}
 \end{equation}
which satisfies Constraint C3 of Problem ${\cal P}_1$. It is readily verified that $\left\{ {{\cal U}^\#,{\bf{w}}^\#} \right\}$ satisfies Constraints C1 and C2 of Problem ${\cal P}_1$. Hence, $\left\{ {{\cal U}^\#,{\bf{w}}^\#} \right\}$ is also feasible for Problem ${\cal P}_1$. On the other hand, if there exists at least one UE whose data rate is strictly larger than its rate requirement, i.e., $r_k^\# > {R_{k,\min }}$. Then, we can adopt the iterative scaling algorithm given in Appendix A of \cite{CunhuTWC} to construct another set of beamforming vectors ${{\bf{w}}^{\# \# }}$ such that $r_k^{\#\#}  = {R_{k,\min }},\forall k$, where $r_k^{\#\#}$ is obtained by substituting ${\bf{w}}^{\#\#}$ into (\ref{datarateUEk}). As a result, $\left\{ {{\cal U}^\#,{\bf{w}}^{\#\#}} \right\}$ is feasible for Problem ${\cal P}_1$.

Based on the above discussions, we arrive at the conclusion that Problem  ${\cal P}_1$ and Problem ${\cal P}_7$ can achieve the same optimal set of selected UEs.

\end{appendices}

\
\






\vspace{-0.8cm}
\bibliographystyle{IEEEtran}
\bibliography{myre}

\begin{thebibliography}{10}
\providecommand{\url}[1]{#1}
\csname url@samestyle\endcsname
\providecommand{\newblock}{\relax}
\providecommand{\bibinfo}[2]{#2}
\providecommand{\BIBentrySTDinterwordspacing}{\spaceskip=0pt\relax}
\providecommand{\BIBentryALTinterwordstretchfactor}{4}
\providecommand{\BIBentryALTinterwordspacing}{\spaceskip=\fontdimen2\font plus
\BIBentryALTinterwordstretchfactor\fontdimen3\font minus
  \fontdimen4\font\relax}
\providecommand{\BIBforeignlanguage}[2]{{%
\expandafter\ifx\csname l@#1\endcsname\relax
\typeout{** WARNING: IEEEtran.bst: No hyphenation pattern has been}%
\typeout{** loaded for the language `#1'. Using the pattern for}%
\typeout{** the default language instead.}%
\else
\language=\csname l@#1\endcsname
\fi
#2}}
\providecommand{\BIBdecl}{\relax}
\BIBdecl

\bibitem{Andrews2014}
J.~Andrews, S.~Buzzi, W.~Choi, S.~Hanly, A.~Lozano, A.~Soong, and J.~Zhang,
  ``What will 5{G} be?'' \emph{IEEE J. Sel. Areas Commun.}, vol.~32, no.~6, pp.
  1065--1082, Jun. 2014.

\bibitem{mugen2016}
M.~Peng, Y.~Sun, X.~Li, Z.~Mao, and C.~Wang, ``Recent advances in cloud radio
  access networks: System architectures, key techniques, and open issues,''
  \emph{IEEE Commun. Surveys Tut.}, vol.~18, no.~3, pp. 2282--2308,
  thirdquarter 2016.

\bibitem{shi2015IWC}
Y.~Shi, J.~Zhang, K.~B. Letaief, B.~Bai, and W.~Chen, ``Large-scale convex
  optimization for ultra-dense cloud-{RAN},'' \emph{IEEE Wireless Commun.
  Mag.}, vol.~22, no.~3, pp. 84--91, Jun. 2015.

\bibitem{Stephen2017}
R.~G. Stephen and R.~Zhang, ``Joint millimeter-wave fronthaul and {OFDMA}
  resource allocation in ultra-dense {CRAN},'' \emph{IEEE Trans. Commun.},
  vol.~65, no.~3, pp. 1411--1423, Mar. 2017.

\bibitem{Yuanming2014}
Y.~Shi, J.~Zhang, and K.~Letaief, ``Group sparse beamforming for green
  {C}loud-{RAN},'' \emph{IEEE Trans. Wireless Commun.}, vol.~13, no.~5, pp.
  2809--2823, May 2014.

\bibitem{dai2016energy}
B.~Dai and W.~Yu, ``Energy efficiency of downlink transmission strategies for
  cloud radio access networks,'' \emph{IEEE J. Sel. Areas Commun.}, vol.~34,
  no.~4, pp. 1037--1050, Apr. 2016.

\bibitem{Binbin2014}
------, ``Sparse beamforming and user-centric clustering for downlink cloud
  radio access network,'' \emph{IEEE Access,}, vol.~2, pp. 1326--1339, Oct.
  2014.

\bibitem{vnhatvt2016}
V.~N. Ha, L.~B. Le, and N.~D. Dao, ``Coordinated multipoint transmission design
  for {Cloud-RANs} with limited fronthaul capacity constraints,'' \emph{IEEE
  Trans. Veh. Technol.}, vol.~65, no.~9, pp. 7432--7447, Sep. 2016.

\bibitem{Abdelnasser2016}
A.~Abdelnasser and E.~Hossain, ``Resource allocation for an {OFDMA}
  {C}loud-{RAN} of small cells underlaying a macrocell,'' \emph{IEEE Trans.
  Mobile Comput.}, vol.~15, no.~11, pp. 2837--2850, Nov. 2016.

\bibitem{pan2017twc}
C.~Pan, H.~Zhu, N.~J. Gomes, and J.~Wang, ``Joint precoding and {RRH} selection
  for user-centric green {MIMO} {C-RAN},'' \emph{IEEE Trans. Wireless Commun.},
  vol.~16, no.~5, pp. 2891--2906, May 2017.

\bibitem{Luong2016}
P.~Luong, L.~N. Tran, C.~Despins, and F.~Gagnon, ``Joint beamforming and remote
  radio head selection in limited fronthaul {C-RAN},'' in \emph{2016 IEEE 84th
  Vehicular Technology Conference (VTC-Fall)}, Sept 2016, pp. 1--6.

\bibitem{Luong2017tsp}
P.~Luong, F.~Gagnon, C.~Despins, and L.~N. Tran, ``Optimal joint remote radio
  head selection and beamforming design for limited fronthaul {C-RAN},''
  \emph{IEEE Trans. Signal Process.}, vol.~65, no.~21, pp. 5605--5620, Nov.
  2017.

\bibitem{Shi2014ICC}
Y.~Shi, J.~Zhang, and K.~B. Letaief, ``{CSI} overhead reduction with stochastic
  beamforming for cloud radio access networks,'' in \emph{2014 IEEE
  International Conference on Communications (ICC)}, 2014, pp. 5154--5159.

\bibitem{Lakshmana2016}
T.~R. Lakshmana, A.~Tolli, R.~Devassy, and T.~Svensson, ``Precoder design with
  incomplete feedback for joint transmission,'' \emph{IEEE Trans. Wireless
  Commun.}, vol.~15, no.~3, pp. 1923--1936, Mar. 2016.

\bibitem{Fan2016}
C.~Fan, Y.~J. Zhang, and X.~Yuan, ``Dynamic nested clustering for parallel
  {PHY}-layer processing in {Cloud-RANs},'' \emph{IEEE Trans. Wireless
  Commun.}, vol.~15, no.~3, pp. 1881--1894, Mar. 2016.

\bibitem{pan2017joint}
C.~Pan, H.~Zhu, N.~J. Gomes, and J.~Wang, ``Joint user selection and energy
  minimization for ultra-dense multi-channel {C-RAN} with incomplete {CSI},''
  \emph{IEEE J. Sel. Areas Commun.}, vol.~35, no.~8, pp. 1809--1824, Aug. 2017.

\bibitem{bai2013evolved}
Z.~Bai, ``Evolved universal terrestrial radio access {(E-UTRA)}; physical layer
  procedures,'' \emph{3GPP, Sophia Antipolis, Technical Specification 36.213 v.
  11.4. 0}, 2013.

\bibitem{Tran2106con}
T.~X. Tran, A.~Hajisami, and D.~Pompili, ``Qua{R}o: A queue-aware robust
  coordinated transmission strategy for downlink {C-RAN}s,'' in \emph{2016 13th
  Annual IEEE International Conference on Sensing, Communication, and
  Networking (SECON)}, June 2016, pp. 1--9.

\bibitem{lau}
V.~K.~N. Lau, F.~Zhang, and Y.~Cui, ``Low complexity delay-constrained
  beamforming for multi-user mimo systems with imperfect csit,'' \emph{IEEE
  Trans. Signal Process.}, vol.~61, no.~16, pp. 4090--4099, Aug. 2013.

\bibitem{Chen2016tvt}
Z.~Chen, X.~Hou, and C.~Yang, ``Training resource allocation for user-centric
  base station cooperation networks,'' \emph{IEEE Trans. Veh. Technol.},
  vol.~65, no.~4, pp. 2729--2735, Apr. 2016.

\bibitem{junzhang2017twc}
J.~Zhang, X.~Yuan, and Y.~J. Zhang, ``Locally orthogonal training design for
  {Cloud-RANs} based on graph coloring,'' \emph{IEEE Trans. Wireless Commun.},
  vol.~16, no.~10, pp. 6426--6437, Oct. 2017.

\bibitem{Nguyen2015}
T.~M. Nguyen and L.~B. Le, ``Joint pilot assignment and resource allocation in
  multicell massive {MIMO} network: Throughput and energy efficiency
  maximization,'' in \emph{2015 IEEE Wireless Communications and Networking
  Conference (WCNC)}, March 2015, pp. 393--398.

\bibitem{CunhuTWC}
C.~Pan, H.~Mehrpouyan, Y.~Liu, M.~Elkashlan, and N.~Arumugam, ``Joint pilot
  allocation and robust transmission design for ultra-dense user-centric {TDD}
  {C-RAN} with imperfect {CSI},'' \emph{IEEE Trans. Wireless Commun.}, vol.~17,
  no.~3, pp. 2038--2053, Mar. 2018.

\bibitem{disu2011}
D.~Su, X.~Hou, and C.~Yang, ``Quantization based on per-cell codebook in
  cooperative multi-cell systems,'' in \emph{2011 IEEE Wireless Communications
  and Networking Conference}, March 2011, pp. 1753--1758.

\bibitem{fangyuantcom}
F.~Yuan and C.~Yang, ``Bit allocation between per-cell codebook and phase
  ambiguity quantization for limited feedback coordinated multi-point
  transmission systems,'' \emph{IEEE Trans. Commun.}, vol.~60, no.~9, pp.
  2546--2559, Sep. 2012.

\bibitem{telatar1999capacity}
E.~Telatar, ``Capacity of multi-antenna {G}aussian channels,'' \emph{European
  transactions on telecommunications}, vol.~10, no.~6, pp. 585--595, 1999.

\bibitem{kailath2000linear}
T.~Kailath, A.~H. Sayed, and B.~Hassibi, \emph{Linear estimation}.\hskip 1em
  plus 0.5em minus 0.4em\relax Prentice Hall Upper Saddle River, NJ, 2000,
  vol.~1.

\bibitem{Jinadl}
N.~Jindal, ``{MIMO} broadcast channels with finite-rate feedback,'' \emph{IEEE
  Trans. Inf. Theory}, vol.~52, no.~11, pp. 5045--5060, Nov. 2006.

\bibitem{Jose2011}
J.~Jose, A.~Ashikhmin, T.~L. Marzetta, and S.~Vishwanath, ``Pilot contamination
  and precoding in multi-cell {TDD} systems,'' \emph{IEEE Trans. Wireless
  Commun.}, vol.~10, no.~8, pp. 2640--2651, Aug. 2011.

\bibitem{Chien2016}
T.~V. Chien, E.~Bjornson, and E.~G. Larsson, ``Joint power allocation and user
  association optimization for massive {MIMO} systems,'' \emph{IEEE Trans.
  Wireless Commun.}, vol.~15, no.~9, pp. 6384--6399, Sep. 2016.

\bibitem{boyd2004convex}
S.~Boyd and L.~Vandenberghe, \emph{Convex optimization}.\hskip 1em plus 0.5em
  minus 0.4em\relax Cambridge university press, 2004.

\bibitem{dinh2010local}
Q.~T. Dinh and M.~Diehl, ``Local convergence of sequential convex programming
  for nonconvex optimization,'' in \emph{Recent Advances in Optimization and
  its Applications in Engineering}.\hskip 1em plus 0.5em minus 0.4em\relax
  Springer, 2010, pp. 93--102.

\bibitem{cunhua2015wcl}
C.~Pan, W.~Xu, W.~Zhang, J.~Wang, H.~Ren, and M.~Chen, ``Weighted sum energy
  efficiency maximization in ad hoc networks,'' \emph{IEEE Wireless Commun.
  Lett.}, vol.~4, no.~3, pp. 233--236, Jun. 2015.

\bibitem{ben2001lectures}
A.~Ben-Tal and A.~Nemirovski, \emph{Lectures on modern convex optimization:
  analysis, algorithms, and engineering applications}.\hskip 1em plus 0.5em
  minus 0.4em\relax SIAM, 2001.

\bibitem{Matskani2008}
E.~Matskani, N.~D. Sidiropoulos, Z.-Q. Luo, and L.~Tassiulas, ``Convex
  approximation techniques for joint multiuser downlink beamforming and
  admission control,'' \emph{IEEE Trans. Wireless Commun.}, vol.~7, no.~7, pp.
  2682--2693, Jul. 2008.

\bibitem{xiaohuge2016}
X.~Ge, S.~Tu, G.~Mao, C.~X. Wang, and T.~Han, ``5{G} ultra-dense cellular
  networks,'' \emph{IEEE Wireless Commun.}, vol.~23, no.~1, pp. 72--79, Feb.
  2016.

\bibitem{access2010further}
E.~U. T.~R. Access, ``Further advancements for {E-UTRA} physical layer
  aspects,'' \emph{3GPP TR 36.814, Tech. Rep.}, 2010.

\bibitem{yeung2007}
C.~K. Au-Yeung and D.~J. Love, ``On the performance of random vector
  quantization limited feedback beamforming in a {MISO} system,'' \emph{IEEE
  Trans. Wireless Commun.}, vol.~6, no.~2, pp. 458--462, Feb. 2007.

\bibitem{zhang2009}
C.~Zhang, W.~Xu, and M.~Chen, ``Robust {MMSE} beamforming for multiuser {MISO}
  systems with limited feedback,'' \emph{IEEE Signal Process. Lett.}, vol.~16,
  no.~7, pp. 588--591, Jul. 2009.

\bibitem{gupta2004handbook}
A.~K. Gupta and S.~Nadarajah, \emph{Handbook of beta distribution and its
  applications}.\hskip 1em plus 0.5em minus 0.4em\relax CRC press, 2004.

\end{thebibliography}


\end{document}